\documentclass[manuscript,screen]{acmart}

\usepackage{subcaption}
\usepackage{longtable}
\usepackage{tabularx}

\AtBeginDocument{%
  }

\setcopyright{rightsretained}
\copyrightyear{2025}
\acmYear{2025}
\acmPrice{}
\acmDOI{10.1145/3742800.3742848}

\acmConference[C\&T 2025]{12th International Conference on Communities \& Technologies}{July 20--23, 2025}{Siegen, Germany}
\acmBooktitle{12th International Conference on Communities \ Technologies (C\&T 2025), July 20--23, 2025, Siegen, Germany}
\acmISBN{979-8-4007-1521-1/25/07}

\begin{document}

%%
%% The "title" command has an optional parameter,
%% allowing the author to define a "short title" to be used in page headers.
% \title[Co-Designing with Multiple Stakeholders and Datasets]{Co-Designing with Multiple Stakeholders and Datasets: A Community-Centered Process to Understand Urban Youth Deviance}

\title[Co-Designing with Multiple Stakeholders and Datasets]{Co-Designing with Multiple Stakeholders and Datasets: A Community-Centered Process to Understand Youth Deviance in the Italian City of Turin}

%%
%% The "author" command and its associated commands are used to define
%% the authors and their affiliations.
%% Of note is the shared affiliation of the first two authors, and the
%% "authornote" and "authornotemark" commands
%% used to denote shared contribution to the research.

\author{Ravinithesh Annapureddy}
%\authornote{Both authors contributed equally to this research.}
\orcid{0000-0003-4383-020X}
\affiliation{%
  \institution{Idiap Research Institute}
  \city{Martigny}
  \country{Switzerland}}
\affiliation{%
  \institution{EPFL}
  \city{Lausanne}
  \country{Switzerland}}
\email{ravinithesh.annapureddy@idiap.ch}

\author{Alessandro Fornaroli}
\orcid{0000-0002-4231-8231}
\affiliation{%
  \institution{Idiap Research Institute}
  \city{Martigny}
  \country{Switzerland}
}
\email{alessandro.fornaroli@alumni.epfl.ch}
\authornote{Former affiliation. The work was conducted while affiliated with the Idiap Research Institute.}

\author{Massimo Fattori}
\orcid{0000-0002-4407-4808}
\affiliation{%
  \institution{Erasmus University Rotterdam}
  \city{Rotterdam}
  \country{Netherlands}
}
\email{fattori@eshcc.eur.nl}

\author{Valeria Lacovara}
\orcid{0009-0001-0050-5215}
\affiliation{%
  \institution{Comune di Torino}
  %\streetaddress{8600 Datapoint Drive}
  \city{Torino}
  \country{Italy}
  %\postcode{78229}
  }
  \email{valeria.lacovara.vl@gmail.com}
  \authornote{Former affiliation. The work was conducted while affiliated with the City of Turin.}

\author{Eleonora Fiori}
\orcid{0009-0002-6384-5863}
\affiliation{%
  \institution{Comune di Torino}
  %\streetaddress{8600 Datapoint Drive}
  \city{Torino}
  \country{Italy}
  %\postcode{78229}
  }
\email{eleonora.fiori@collaboratori.comune.torino.it}

\author{Sarah Vollmer}
\orcid{0009-0009-7889-1855}
\affiliation{%
  \institution{Camino}
  \city{Berlin}
  \country{Germany}}
\email{sarahvollmer@camino-werkstatt.de}

\author{Moritz Konradi}
\orcid{0009-0007-7743-4033}
\affiliation{%
  \institution{Camino}
  \city{Berlin}
  \country{Germany}}
\email{moritzkonradi@camino-werkstatt.de}

\author{Britta Elena Hecking}
\orcid{0009-0009-8071-8947}
\affiliation{%
  \institution{Camino}
  \city{Berlin}
  \country{Germany}}
\email{brittahecking@camino-werkstatt.de}

\author{Gianfranco Todesco}
\orcid{0009-0000-5831-6209}
\affiliation{%
  \institution{Comune di Torino}
  %\streetaddress{8600 Datapoint Drive}
  \city{Torino}
  \country{Italy}
  %\postcode{78229}
  }
\email{gianfranco.todesco@collaboratori.comune.torino.it}

\author{Daniel Gatica-Perez}
\orcid{0000-0001-5488-2182}
\affiliation{%
  \institution{Idiap Research Institute}
  \city{Martigny}
  \country{Switzerland}}
\affiliation{%
  \institution{EPFL}
  \city{Lausanne}
  \country{Switzerland}}
\email{gatica@idiap.ch}

%%
%% By default, the full list of authors will be used in the page
%% headers. Often, this list is too long, and will overlap
%% other information printed in the page headers. This command allows
%% the author to define a more concise list
%% of authors' names for this purpose.
\renewcommand{\shortauthors}{Annapureddy et al.}

%%
%% The abstract is a short summary of the work to be presented in the
%% article.
\begin{abstract}

  This paper presents the co-design and design evaluation of \textit{Sbocciamo Torino} civic tool, which helps understand and act upon the issues of youth deviance in the Italian city of Turin through multi-stakeholder collaboration and collaborative data analysis. Rooted in research through design and participatory design methodologies, the civic tool integrates a data dashboard, stakeholder committee, and structured co-design sessions to facilitate collaborative analysis and intervention planning. The civic tool was developed in partnership with municipal authorities, law enforcement, NGOs, and social services, and reflects their institutional priorities while centering community knowledge. We describe the iterative co-design process, including stakeholder workshops for design, validation, training, and evaluation. The civic tool's impact on stakeholder trust, collaboration, and decision-making was assessed through surveys and open-ended questionnaires. Our findings show that stakeholders valued the inclusive design approach and data-driven collaboration while revealing barriers in communication, data literacy, and operational coordination. Furthermore, political and institutional support was identified as critical to the civic tool's success. This paper contributes to research on community technologies by demonstrating how civic tools can be collaboratively developed to navigate wicked social problems through participatory design. 
  
\end{abstract}

%%
%% Keywords. The author(s) should pick words that accurately describe
%% the work being presented. Separate the keywords with commas.

\received{31 March 2025}
\received[revised]{05 June 2025}
% \received[accepted]{5 June 2009}

%%
%% This command processes the author and affiliation and title
%% information and builds the first part of the formatted document.
\maketitle

\section{Introduction}

Urban communities today face increasingly complex and localized challenges that intersect social, economic, and environmental concerns. These issues --- ranging from housing insecurity to youth violence~\cite{un2022envisaging} --- are often described as “wicked problems”~\cite{Rittel1973}, characterized by their resistance to straightforward solutions, evolving nature, and deeply embedded contextual factors. Traditional civic decision-making processes, typically structured around hierarchical institutions and compartmentalized responsibilities, often fall short of addressing these complexities. As a result, the voices of affected communities may go unheard, and responses risk being ineffective or misaligned with local realities~\cite{Bryson2013DPPP}.
To engage with such problems more effectively, scholars and practitioners have increasingly advocated for models of \textit{collaborative governance}~\cite{Emerson2011ColabGov}. These models emphasize the need for inclusive, cross-sectoral collaboration, recognizing that no single agency or organization can fully understand or resolve wicked problems in isolation. 

At the same time, digital transformation is reshaping the way cities govern. Alongside investments in physical infrastructure, cities are increasingly adopting Information and Communication Technologies (ICT) to enhance administrative capabilities and enable more proactive, data-informed governance~\cite{chourabi2012, pereira2018smart}. ICT systems can support collaboration by enabling new forms of data collection, visualization, and participatory engagement~\cite{Yuan2019coprod}. Prior work has explored how technology can facilitate local decision-making and empower communities through data-supported civic participation~\cite{Saldivar2019, Kuznetsov2011, ledantec2012}.
Yet, the promise of participatory, data-driven governance is not easily realized in practice. Despite widespread recognition of the need for collaboration, institutional and organizational disparities --- such as uneven digital skills, limited data literacy, and misaligned working cultures --- often hinder meaningful stakeholder engagement~\cite{Merkel2004, Taylor2013, Gilman2017, Boehner2016, Stoll2012}. Without addressing these asymmetries, participatory processes risk reproducing existing power imbalances or becoming performative.

In this paper, we consider the concept of \textit{civic tools} as a way to address these challenges. We define civic tools as resources --- including human, physical, and digital components --- that support meaningful engagement between stakeholders and decision-makers. Unlike traditional civic technologies that often prioritize technological innovation or open data access~\cite{Saldivar2019}, civic tools foreground human-centered design and procedural collaboration. 
Designing such civic tools for data-assisted, multi-stakeholder decision-making raises several interrelated challenges. First, aligning stakeholder priorities and building mutual trust requires ongoing coordination and facilitation. Second, disparities in data literacy and technological capacity can limit inclusive participation and skew decision-making. Third, fostering sustained collaboration demands carefully designed mechanisms for engagement, co-design, and shared deliberation.

In this work, we address these challenges by focusing on youth deviance --- a term we use to describe a spectrum of behaviors by young people that range from socially disruptive acts to activities considered unlawful, depending on the local context~\cite{Kim2017}. While some of these behaviors may fall under the jurisdiction of juvenile justice systems, many others are managed informally by schools, social services, or municipal programs. Youth deviance constitutes a wicked problem: it emerges from complex social, economic, and cultural dynamics, and resists simple or uniform solutions~\cite{Dodge2006}. Traditionally, institutional responses have relied on top-down, punitive approaches. However, growing evidence suggests that effective interventions must be rooted in a nuanced understanding of local conditions and the structural factors that shape youth behavior~\cite{Crawford2023, Dodge2006}. Addressing such issues requires coordinated action among diverse actors, including city administrations, law enforcement, social services, educators, and non-governmental organizations (NGOs)~\cite{Unlu2021}.

We investigate these challenges in the context of youth deviance prevention in the Italian city of Turin, which in recent years has faced increasing concerns about spontaneous and sometimes violent youth group formations. Despite the existence of various institutional mechanisms, local authorities --- including police and municipal offices --- have struggled to develop coordinated, early-stage responses to these issues. A lack of shared understanding across agencies and limited access to contextualized data were identified as key barriers.
To address this gap, we co-developed \textit{Sbocciamo Torino} to support multi-stakeholder collaboration around youth deviance prevention. The civic tool aims to enable structured data sharing, joint analysis, and participatory decision-making among a diverse network of actors. It was developed in collaboration with the city administration as part of a large European project.
This paper is guided by the following research questions:

   \begin{enumerate}
        \item  \begin{quote} \textbf{RQ1:} How can participatory civic tools be co-designed to facilitate data-informed, multi-stakeholder collaboration in decision-making to support youth deviance prevention in the context of Turin? \end{quote}
        \item  \begin{quote}  \textbf{RQ2:} Which practical considerations of communities shape the design of such civic tool, given the specific conditions of the city, including institutional processes and stakeholder attitudes?  \end{quote}
    \end{enumerate}

We adopted a \textit{Research through Design} (RtD) approach~\cite{Zimmerman2007, Gaver2012} to iteratively co-design the tool with stakeholders from municipal offices, local police, social services, and non-governmental organizations (NGOs). Through this process, \textit{Sbocciamo Torino} evolved as both a technological platform and a procedural infrastructure that reconfigures how stakeholders interact, share data, and deliberate.
The resulting civic tool consists of three core components: \textbf{(1)} a committee of relevant stakeholders responsible for agenda-setting and coordination, \textbf{(2)} a data dashboard that centralizes and visualizes evidence to support decision-making, and \textbf{(3)} a structured mechanism for regular meetings, collaborative data analysis, and co-design sessions. Together, these components constitute a governance network aimed at supporting early intervention strategies through shared understanding and collective action.

The design of \textit{Sbocciamo Torino} builds on a long-standing interest in HCI in creating infrastructures that support collaboration across institutions, disciplines, and communities. Prior work has shown how digital technologies can enable new forms of participation, but also how power imbalances, differences in expertise, and institutional constraints can undermine collaborative intentions~\cite{Uslaner2005, Harrington2019, Leonardi2023}. In response, our work focuses on how participatory civic tools can be designed not only to enable data use but also to support inclusive and sustained forms of decision-making in public governance.
Specifically, our work makes the following contributions:

\begin{itemize}
    \item \textbf{A design approach for civic collaboration:} We introduce a structured method for co-designing civic tools that support data-informed collaboration among diverse stakeholders, tailored to the institutional realities and practical needs of city administrations.
    
    \item \textbf{A process model for participation across differences:} We describe how the participatory process was organized to account for differences in roles, capacities, and data literacies, emphasizing the conditions and practices that made participation meaningful and constructive.
    
    \item \textbf{A situated case of civic tool development:} We provide a richly contextualized account of how a civic tool was developed in response to a real-world urban challenge, contributing practical insights for others designing collaborative tools in complex public sector settings.
\end{itemize}

While this paper focuses on the co-design process and the resulting structure of the civic tool, we do not assess the long-term outcomes of the interventions it may support. The remainder of the paper is organized as follows: Section~\ref{sec:related-work} reviews related literature on participatory civic technologies and collaborative governance. Section~\ref{sec:methods} outlines our methodology, including the Research through Design approach and participatory process. Section~\ref{sec:pd-outcomes} presents the outcomes of the co-design activities. Section~\ref{sec:components} describes the civic tool’s components. Section~\ref{sec:results} discusses the development process and lessons learned. Finally, Sections~\ref{sec:discussion} and~\ref{sec:conclusion} reflect on the implications of this work for future civic tool design and participatory practices.

\section{Related Work} \label{sec:related-work}

Our work is related to several bodies of literature, spanning criminology, group deliberation, collaborative governance, and participatory design. Each of these threads of previous works is concisely discussed in this section.

\subsection{Criminology Literature on Youth Deviance}

Youth deviance and its widely known form, juvenile delinquency, have been extensively studied, and preventive measures have significantly advanced in recent years~\cite{Sullivan2010, Gupta2022}. There is a lack of a universal definition of juvenile delinquency, as its meaning can vary by context, region, culture, and application~\cite{Crawford2023}. For this paper, we consider this to be an umbrella term for all behaviors ranging from antisocial to criminal that can put young people at risk~\cite{Kim2017}. Criminological research identifies multiple interdependent factors, such as social surroundings, personal growth, loss of hope for the future, and feelings of marginalization, that influence young individuals to engage in deviant behavior~\cite{Boehnke2002, Vazsonyi2018}. Additionally, economic hardships within families, high unemployment rates, inadequate educational opportunities, school dropouts, and identity struggles have consistently emerged as primary causes of spontaneous and sometimes violent behavior among youth~\cite{Hirschfield2011, Murray2010, Loeber2009}. Addressing such a wicked problem effectively requires comprehensive strategies informed by evidence-based insights and multi-stakeholder collaboration. Despite existing systems and networks, local authorities lack sufficient insight into group dynamics and a comprehensive understanding of this deviance phenomenon~\cite{Crawford2023}. Given the vulnerability of young people, local authorities must formulate comprehensive policies through partnerships to promote inclusion and prevent social, economic, and political marginalization~\cite{Crawford2023}. The complexity and variability of youth deviance demand coordinated efforts among multiple stakeholders, including educational institutions, social services, law enforcement, community organizations, families, and youth themselves~\cite{Howell2009}. Such collaborations are essential to ensure that preventive and responsive measures reflect the nuanced reality of youth experiences and sociocultural contexts. In addition, the participation and consultation of stakeholders are particularly relevant as their collaboration is often required for implementing the measures~\cite{Crawford2023}. Our work attempts to provide a multi-stakeholder collaborative governance framework supported by multi-stakeholder data to provide a shared understanding of the multifaceted factors driving youth deviance. This approach not only promotes more inclusive and contextually relevant strategies but also enhances the effectiveness and legitimacy of local policies and interventions~\cite{Crawford2023}. 

\subsection{Collaborative Governance}

Collaborative Governance (CG) offers a framework that enables cooperation among various sectors, including local governments, law enforcement, educational institutions, social services, NGOs, and community organizations~\cite{Ansell2007}. This approach emphasizes shared decision-making, resource pooling, and mutual accountability, ensuring that interventions are more holistic and sustainable than traditional top-down approaches~\cite{gollagher2013}. It differs from e-governance in providing inclusive decision-making, rather than improving services~\cite{Gilman2017}. By bringing together stakeholders with different types of data and experiences, CG allows for a more nuanced understanding of wicked problems~\cite{Weymouth2015}. The three key elements of a CG are stakeholder engagement, collective motivation, and the ability to take coordinated action. These elements facilitate data integration and encourage collective responsibility, and trust for sharing information while collaborating toward common objectives~\cite{Bryson2015}. The need for collaborative strategies that engage various stakeholders in the community and promote evidence-informed solutions is widely recognized in community technology research~\cite{Saldivar2019}. Recent studies explored the development of such platforms that enable multi-stakeholder collaboration, focusing on approach, design, motivations, and frameworks~\cite{Carroll2013, Ansell2017, Crivellaro2019, Balestrini2017, Brandon2021opencivic}. Diversity, open dialogue, continuous learning, and cooperative problem-solving formed support for these platforms. They also underscore the importance of mediation in handling complex collaborative relationships. However, these same components can make implementing CG challenging. Therefore, it is necessary to develop structured CG models that provide a structured environment, whether physical or digital, to enable the exchange of diverse skills, resources, knowledge, and needs~\cite{Ansell2017}. There is limited work on how to co-design participatory CG structures, and our work attempts to develop one such model to operationalize collaborative governance in urban decision-making.

\subsection{Participatory Design in Civic Contexts}

Participatory design (PD) is a method that actively involves end-users and stakeholders in the design process to ensure that the resulting technologies meet their real-world needs~\cite{Merkel2004, Bratteteig2016-1}. PD has become a core principle in many of the works on community and technology interaction. In recent years, it has been widely adopted in developing civic technologies and algorithms to improve inclusivity, transparency, and user-centeredness~\cite{Bodker2015, Slingerland2019, Wulf2011, Freeman2017, Harrington2019, Saxena2020, flugge2021, Leonardi2023, Bohoj2011}. This approach is particularly valuable for civic tools, where systems must address complex social issues and interact with multiple sectors. In these works, community members, policymakers, and researchers work together to identify challenges, co-create solutions, and iteratively refine the technologies being developed. However, existing HCI research has not fully explored participatory design in public safety and youth deviance. Prior studies are focused on the use of design thinking and its variants, like crime prevention through environmental design~\cite{Davey2005, Cozens2015}, applied to safety and crime-related challenges in urban settings. However, to our knowledge, there is limited research on integrating participatory civic tools with real-time stakeholder decision-making. Our work addresses this gap by co-designing a participatory civic tool that engages local communities in addressing youth deviance.

Nevertheless, researchers have identified several limitations of the PD approach~\cite{Kensing1998}. First, conducting PD can be expensive and time-consuming compared to other design methods~\cite{Spinuzzi2005, Robertson2012}. Second, PD methods originated in small and local settings; when applied to complex public system settings with multiple stakeholders, they encounter considerable challenges~\cite{Zahlsen2023}. Third, the simple involvement of end users does not automatically yield a better solution, if any solution~\cite{ledantec2012}. The unpredictability and involvement of less experienced participants means that the results can vary, and extra effort is needed to constantly revise the next steps and translate the insights into actionable challenges~\cite{Bratteteig2016-1}. Fourth, in many PD projects, end-users contribute ideas but do not wield final decision-making power in the design process~\cite{Volkmann2023}. This can lead to power imbalances and cultural factors associated with participation. These long-term time commitments and resource demands can strain the PD projects.

Previous studies have focused on cooperative and data-driven decision-making systems tailored for community engagement and collective deliberation. These prior studies focus on facilitating online public discussion on policy~\cite{Bohoj2011, Kriplean2012}, supporting citizen science initiatives~\cite{Luther2009}, and information visualization for consensus building and decision making~\cite{Mahyar2017, Moghaddam2015, Jin2017}. These works have shown that these technologies can move users from passive data collectors to actively analyzing the data~\cite{Luther2009}, and designing lightweight, scalable solutions can already support deliberations compared with high-investment and complex platforms~\cite{Kriplean2012}. They further posit that visual cues and lightweight moderation constraints significantly aid users toward group consensus~\cite{Mahyar2017, Jin2017, Moghaddam2015} but also are examples of bringing personal experiences together with the data visualizations. These studies illustrated how technology can cultivate confidence by sharing information and strengthen and inform the group's decision-making ability. However, most of the studies involved either homogeneous groups --- where participants belong to the same community or professional background --- or individual citizens engaging independently in deliberative processes. Recent research also underscored the necessity of structured support from experienced to less experienced group members, advocating the provision of relevant contextual information to help participants interpret complex data effectively~\cite{Boehner2016, Puussaar2018}. The role of technical infrastructure in facilitating equitable participation has also received attention, highlighting the importance of inclusive design in collaborative tools~\cite{Lindley2017}. Significant attention has been given to participatory approaches, wherein stakeholders actively partake in designing civic technologies~\cite{Boehner2016, Choi2017, Leonardi2023}. Such research emphasizes several key factors crucial for successful deliberation: interdisciplinarity; data literacy; using the skills of individuals to improve cooperation within diverse groups; using collaboration as a method for data analysis; and reliance on synchronous communication. Our work extends this research by looking into a different user group, who are stakeholders representing different organizations and data-assisted decision-making with multiple stakeholders.

\section{Methods}\label{sec:methods}

\subsection{Research through Design}

This work follows a `Research through Design' methodology~\cite{Zimmerman2007}, employing a participatory design approach. This methodology enabled co-design of the civic tool based on stakeholder needs, iterative refinements, and validation through real-world use~\cite{Gaver2012}. Research through Design is particularly well-suited for participatory civic tools, as it emphasizes the creation of artifacts as a means of knowledge generation and enables iterative prototyping and reflective engagement with stakeholders~\cite{Zimmerman2014}. In addition, we adapted the Participatory Design (PD) process to develop \textit{Sbocciamo Torino} civic tool in Turin. Unlike traditional technology development approaches where stakeholder input is limited to feedback or usability testing, PD ensures that stakeholders are embedded throughout the entire design and decision-making process~\cite{Spinuzzi2005, Bannon2018}. We held multiple workshops and consultations with stakeholders to refine and validate the design with representatives from municipal services, NGOs, local police, and community organizations. The process was iterative and flexible, allowing adjustments at different stages. The city administration coordinated and convened the local stakeholders. Design decisions emerged through open collaboration between researchers, administrators, and diverse stakeholder groups during the PD process.
Due to the iterative nature of the co-design process, several activities that began as methodological steps are reported later as results, because their outcomes informed subsequent stages.

\subsection{Design Steps of \textit{Sbocciamo Torino}}

%TC:ignore
\begin{figure}[ht]
  \centering
\includegraphics[width=\linewidth]{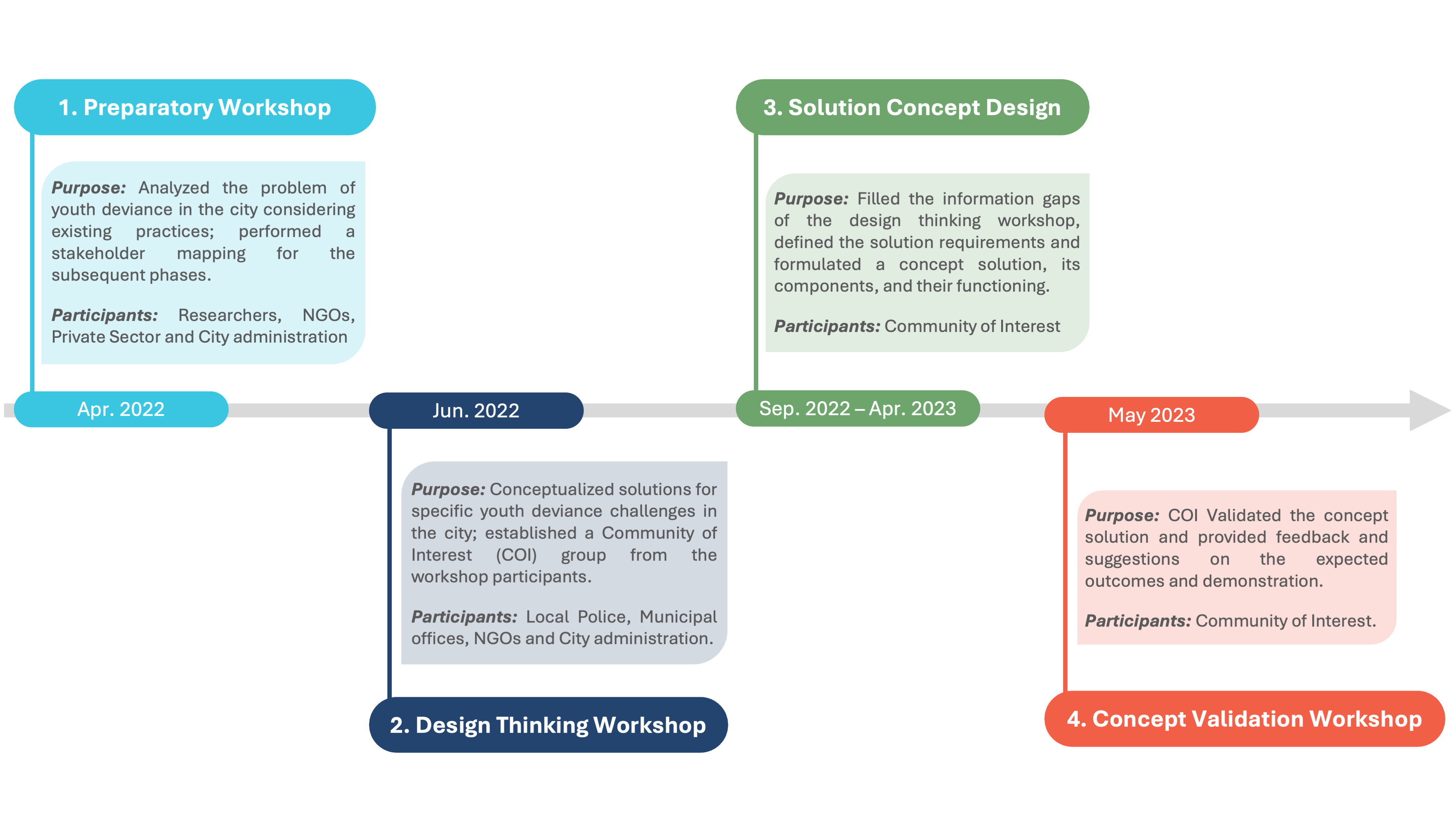}
  \caption{Process steps of the participatory design methodology} \label{fig:design-steps}
  \Description{}
\end{figure}
%TC:endignore

The process began with a preparatory workshop (Step 1), which identified stakeholders concerned with addressing youth deviance in Turin. These stakeholders (Table~\ref{tab:participants-design-thinking}) participated in a Design Thinking workshop (Step 2) to align local needs with priorities identified in a literature review on youth deviance.
A Community of Interest (COI) subgroup (Table~\ref{tab:participants-concept-validation}) was formed from the participants to provide input throughout \textit{Sbocciamo Torino}'s design and development.
% A subgroup of participants (Table~\ref{tab:participants-concept-validation}) referred to as the Community of Interest (COI) in this paper was formed to provide ongoing input throughout \textit{Sbocciamo Torino}'s design and development.
After addressing the design thinking workshop gaps, we proposed a solution concept (Step 3) and its components, which were validated by the COI (Step 4). Then, we created a final design incorporating the COI feedback. Figure~\ref{fig:design-steps} summarizes the steps and participants. Our study received necessary approvals from the research institute's ethical review board, and written consent was obtained from participants.

\subsection{Development and Demonstration Steps}

\begin{figure}[ht]
  \centering
\includegraphics[width=\linewidth]{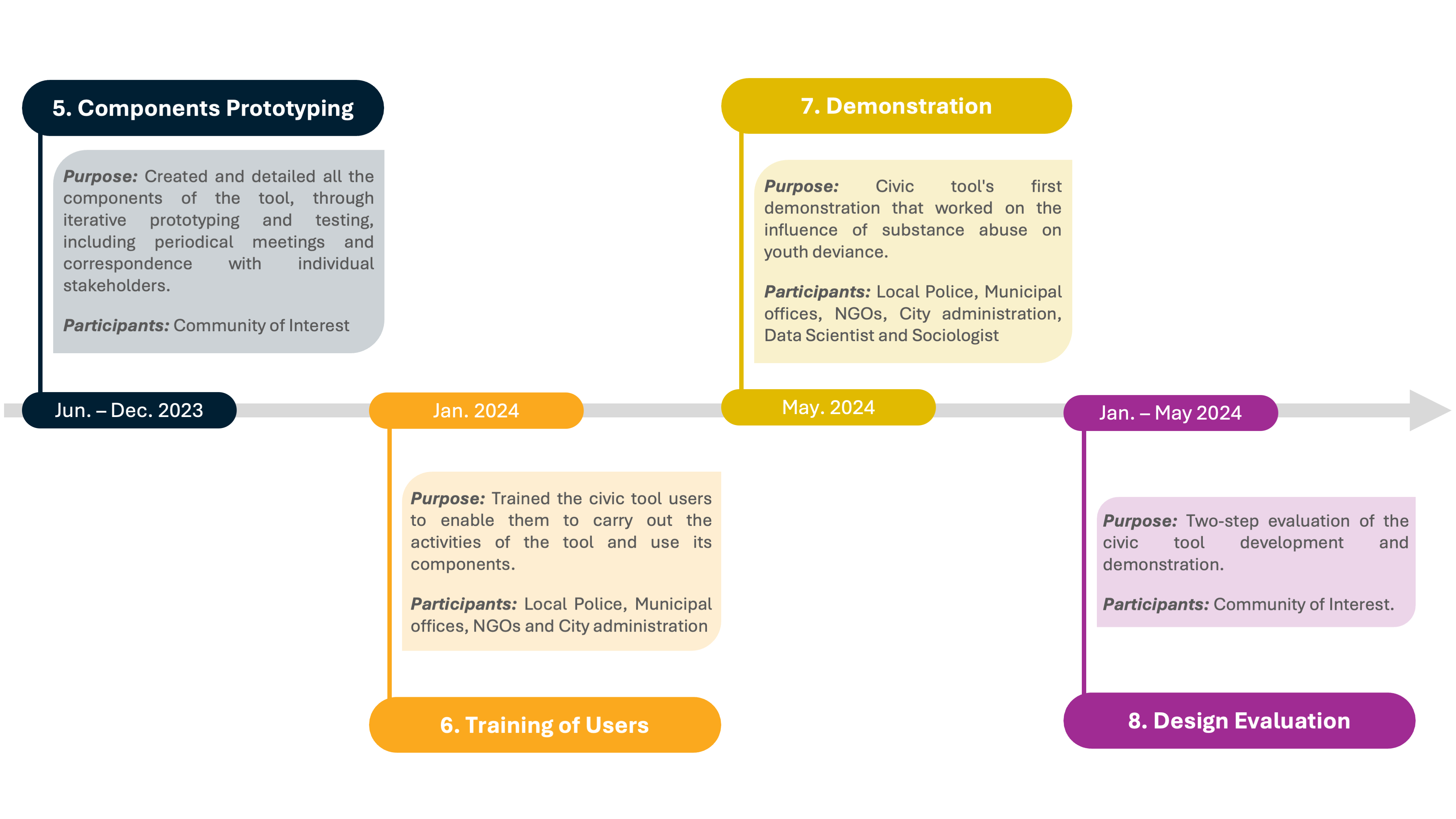}
  \caption{Steps followed during the development and testing phase} \label{fig:implementation-steps}
  \Description{}
\end{figure}

During the development phase, we designed and prototyped components of the civic tool (Step 5) together with the COI.
% An overview of the development and demonstration steps is presented in Figure~\ref{fig:implementation-steps}.
Figure~\ref{fig:implementation-steps} overviews the development and demonstration steps.
The demonstration phase began with training the end-users (Step 6). While most of these users had been involved in the prototyping, not all of them had interacted with every component. 
All users attended training to establish a common understanding.
% To ensure effective use, all users went through a training session to be able to operate from a common knowledge of the civic tool.
Then, we demonstrated the tool (Step 7) where the stakeholders worked on the agenda of the intersection of youth deviance and drug abuse. 

%TC:ignore
%TC:endignore

\subsection{Design Evaluation}\label{sec:design-evaluation}

The process concluded with a design evaluation of the civic tool (Step 8, Section~\ref{sec:evaluation}, Figure~\ref{fig:implementation-steps}). The evaluation of \textit{Sbocciamo Torino} was structured around a logic model~\cite{Paul2001} that mapped the tool’s intended impact, outcomes, and mechanisms. This approach ensured that evaluation criteria were linked to the civic tool’s goals and remained relevant to stakeholder needs~\cite{Bossen2016}.
The evaluation was embedded in the training and demonstration steps using a mixed-method approach.
% Given the nature of the participatory process, this evaluation was embedded into the civic tool’s training and demonstration steps that used a mixed-method approach.
First, a survey was administered to the users (Table~\ref{tab:part1-eval-users}) after the training session. Second was a post-training survey, detailed in Appendix~\ref{sec:appendix-survey-january}, which included seven Likert scale questions and one question asking participants to identify their organization. Later, an open-ended questionnaire (provided in Appendix~\ref{sec:appenix-questionnaire-may}) was given to participants during the demonstration session. 
The surveys were designed to assess understanding, perceived usefulness, inclusiveness, and areas for improvement from multiple stakeholder perspectives.
We analyzed the quantitative responses using descriptive statistics to evaluate general trends in participants’ comprehension, attitudes toward inclusivity, and confidence in using the tool. 
The responses to the open-ended questions were analyzed thematically using open coding to extract common themes and insights.
The difficulty in involving all users throughout the process did not permit us to conduct one-to-one interviews with them. Therefore, the design evaluation was limited to a survey and a questionnaire.

\section{Outcomes of the Participatory Design Process (RQ1)}\label{sec:pd-outcomes}

\subsection*{Step 1: Preparatory Workshop}

%TC:ignore
\begin{table}[ht]
  \caption{Participants to the Preparatory Workshop} \label{tab:participants-preparatory}
  \begin{tabular}{p{0.015\linewidth}p{0.02\linewidth}p{0.47\linewidth}p{0.4\linewidth}}
    \toprule
    \textbf{No.} & \textbf{Sex}  & \textbf{Organization type} & \textbf{Position} \\
    \midrule
    1 & F  & Turin city administration  & Project manager \\
    2 & M  & Turin city administration  & Chief commissioner \\
    3 & M  & A Polish city's municipal police  & Chief commandant \\
    4 & M  & International network organization on restorative justice  & Chair of the Board \\
    5 & M  & University & Urban security processes researcher \\
    6 & M  & University  & Media and communication researcher \\
    7 & F  & University & Criminology researcher \\ \bottomrule
\end{tabular}
\end{table}
%TC:endignore

To discuss the youth deviance challenges in Turin, we conducted a preparatory workshop with representatives from local police, researchers, and practitioners in criminology, media studies, urban safety, and HCI (Participants in Table~\ref{tab:participants-preparatory}). The workshop focused on reflecting on previous practices and experiences in addressing similar issues and identifying stakeholders for the next steps of the design process.

% The city of Turin was already implementing a range of measures, from tackling educational poverty and promoting spaces of aggregation to a restorative justice approach to achieve social cohesion through community involvement.
Turin was already implementing educational, restorative-justice, and community-cohesion measures. The analysis of experiences\footnote{The 9th author has practical, on-the-ground knowledge of these issues, gained through experience working in the local police in Turin.} indicated that previous measures often failed because the dynamics of these spontaneous aggregations of violent groups were not known, and there is hardly any reliable data~\cite{Crawford2023}. The first part of the workshop provided a better understanding of this complex phenomenon in the city to the researchers and an overview of existing prevention practices across the world to the city officials. 

The second part of the preparatory workshop involved identifying the relevant stakeholders related to youth deviance in Turin. Stakeholder mapping (Figure~\ref{fig:stakeholder-mapping} in Appendix~\ref{app:stakeholder-mapping}) showed that a wide variety of actors were exerting influence at different levels of prevention. The main stakeholders identified for the primary and secondary levels of prevention are: Regional Administration, City Administration, NGOs, and Street and Community Educators, and at the tertiary level are: Local Police and Juvenile Prosecutor’s Office. The city then invited these stakeholders for a Design Thinking workshop (Step 2, Section~\ref{sec:dt-workshop}).
% to brainstorm further on the first outcome of the participatory workshop.

\subsection*{Step 2: Design Thinking Workshop}\label{sec:dt-workshop}

%TC:ignore
\begin{table}[ht]
  \caption{Participants to the Design Thinking Workshop} \label{tab:participants-design-thinking}
  \begin{tabular}{p{0.015\linewidth}p{0.7\linewidth}p{0.18\linewidth}}
    \toprule
    \textbf{No.} & \textbf{Organization} & \textbf{Number of Participants (Sex)} \\
    \midrule
    
    1 & NGO promoting territorial education, intercultural engagement, and community building among adolescents and young people & 1 (M=1) \\
    
    2 & NGO promoting programs that foster inclusion, youth participation, and the social well-being of minors and adults & 2 (M=1, F=1) \\
    
    3 & NGO's project for prevention, rehabilitation, inclusion, and growth for minors of foreign origin at risk of deviance & 2 (M=1, F=1) \\

    4 & NGO supporting individuals facing social marginalization, addiction, and isolation & 1 (F=1) \\
    
    5 & NGO promoting civic participation of young people in democratic life & 2 (M=2) \\
    
    6 & NGO promoting sport as a right for all citizens  & 1 (M=1) \\
    
    7 & NGO promoting use of evaluation in public administrations, with a focus on impact assessment & 1 (F=1) \\
    
    8 & Local Health services (Addictions department) & 2 (F=2) \\
    
    9 & Office of the prisoners’ ombudsperson & 1 (F=1) \\
    
    10 & City's Educational services (School inclusion department) & 2 (M=1, F=1) \\
    
    11 & Local police (Community policing) & 2 (M=1, F=1) \\

    12 & Local police (other units) & 2 (M=1, F=1) \\

    13 & Local police (Juvenile Court) & 2 (F=2) \\

    14 & Journalist & 1 (M=1) \\

  \bottomrule
\end{tabular}
\end{table}
%TC:endignore

The Design Thinking workshop initiated the consultation process with local stakeholders, including police officers, NGO members, journalists, social workers, city officials, representatives from the office of school inclusion, the prisoners' ombudsperson's office, and the juvenile court. Twenty-two participants (12 Female and 10 Male) attended (See Table~\ref{tab:participants-design-thinking}) attended the workshop. Most of them were involved in other city initiatives related to youth deviance. The fourth author facilitated the workshop in Italian, and the results were translated into English. Participants, divided into four groups, identified three challenges related to prevention measures: (1) Addressing the needs and rights of minors and vulnerable populations; (2) Implementing continuous, structured programs by the city rather than irregular, short-term projects and (3) the difficulty of sustainably delivering municipal services, which currently fall short in tackling complex urban issues, particularly in supporting families, involving schools, and fostering collaboration between services.
The most pressing challenge identified was the disconnect between the short-term focus of political decision-making and the need for long-term strategies.

Groups have proposed various solutions, including (1) Strengthening community dialogue through mediation and collaboration with street educators; (2) Establishing a permanent working group to facilitate dialogue between local actors and policymakers; (3) Creating ``neighborhood houses'' by rehabilitating unused urban spaces and spaces for systematic youth participation; (4) Launching communication campaigns, supported by influencers, to promote positive behaviors and (5) Developing a mobile intervention model based on street work to engage target groups in informal settings.
However, no actionable recommendations for action came out of the workshop itself. Nonetheless, it was clear that there was a gap between the politicians and stakeholders in the social sector.

\subsection*{Step 3: Solution Concept Design}\label{sec:initial-design}

Based on the findings from the design thinking workshop, we engaged with the city administration to narrow down to a specific goal and we formulated the core goal for the solution as: ``\textit{In what ways might we support collaborative decision-making to tackle youth deviance and enable evidence-based intervention?}'' and defined the tool requirements. Addressing this question required the collaboration of diverse stakeholders, each with access to specific data that could inform decision-making. This approach supports the design of more effective prevention policies tailored to different problem contexts and circumstances~\cite{Ansell2007, Howell2009}.

Based on these, we proposed a conceptual design of the solution (Figure~\ref{fig:ST-functioning}). The concept was to constitute a civic tool based on collaborative data analysis among stakeholders to inform the policy-making process. 

The concept comprised three essential components:
\begin{enumerate}
    \item A \textbf{committee of stakeholders} from various organizations involved in preventing the problem of youth deviance in the city.
    
    \item A \textbf{data dashboard} featuring visualizations of relevant data, sourced from public sources and databases owned by the stakeholders.
    
    \item \textbf{Periodical meetings} of the committee of stakeholders, where collaborative data analysis would take place, leading to the drafting of policy ideas.
\end{enumerate}

\subsection*{Step 4: Concept Validation Workshop}\label{sec:concept-validation-workshop}

%TC:ignore
\begin{table}[ht]
\caption{Participants to the Concept Validation Workshop}
\label{tab:participants-concept-validation}
  \begin{tabular}{ccc}
    \toprule
    \textbf{ID} & \textbf{Type of Organization} & \textbf{Number of Participants} \\
    \midrule
     1 & Third Sector (NGO, association, cooperative) & 3 \\ \\

     2 & City's Educational Services Department  & 1 \\ \\

     2 & City's Prisoners' Ombudsperson's Office & 1 \\ \\

     3 & Local Police (from various units) & 4 \\

  \bottomrule
\end{tabular}
\end{table}
%TC:endignore

A concept validation workshop was held with the Community of Interest (COI) (Table~\ref{tab:participants-concept-validation}) to gather feedback on the conceptual solution and adjust its components before developing them. Following a semi-formal focus group method, we presented the tool's components for data processing, visual data analysis, co-decision processes, and assessment.

% The COI validated the concept, expressing 
During the focus group, the COI expressed that the tool could consolidate distributed data on youth deviance, facilitate precise actions, incorporate multiple perspectives, and track progress. During the concept validation workshop, it came to light that some NGOs lacked organized data collection and maintenance methods. To address this, the design was adjusted to streamline these processes, providing support for data visualization and integrating it into decision-making. Specifically, a project manager, a data scientist, and a sociologist were made part of the civic tool to support the stakeholders. 
In addition, there was also a discussion about the frequency of committee meetings. They all deliberated and agreed to have the meetings every six months or as requested by the majority of the committee members.

\subsection*{Step 5: Components Development and Prototyping}

%TC:ignore
\begin{figure}[ht]
  \centering
\includegraphics[width=0.75\textwidth]{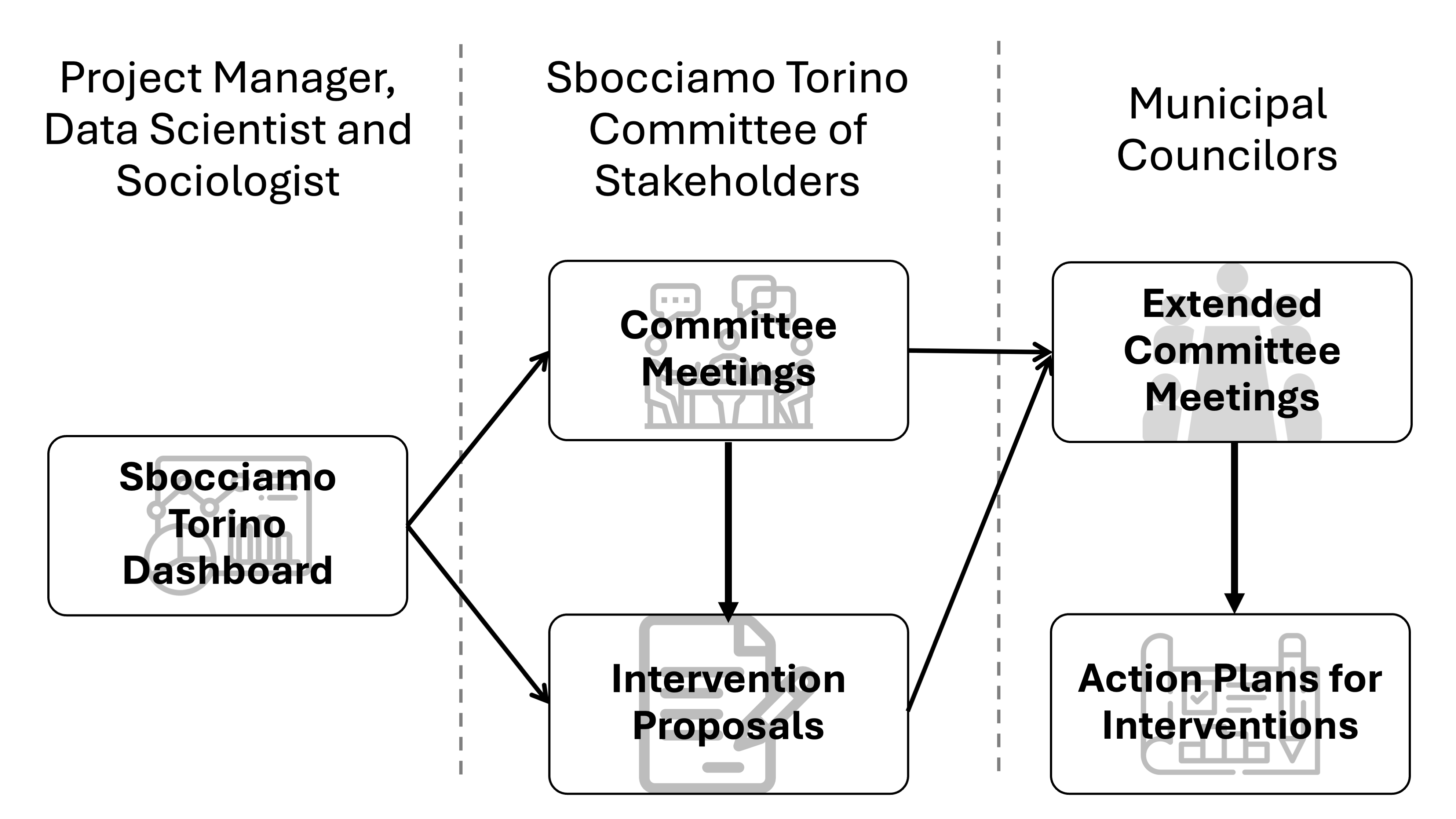}
  \caption{Functioning of \textit{Sbocciamo Torino}} \label{fig:ST-functioning}
  \Description{Process Flow of \textit{Sbocciamo Torino}. This diagram outlines the workflow involving the project manager, data scientist, sociologist, the committee of stakeholders, and municipal councilors. The process begins with data inputs from \textit{Sbocciamo Torino} Dashboard, which are discussed in committee meetings to develop intervention proposals. These proposals are then reviewed in extended committee meetings with municipal councilors, resulting in actionable intervention plans. The flow emphasizes collaboration across different roles and iterative feedback to refine intervention strategies.}
\end{figure}
%TC:endignore
 
In this step, the solution concept was developed into a complete civic tool design --- \textit{Sbocciamo Torino}. 
Throughout the development process, COI members were consulted through individual meetings and correspondence to obtain their feedback on the prototypes of the components of the civic tool. Prototyping activities included gathering feedback on the organization of the meetings of the committee members; the data to be included in the dashboards; interactions with the dashboard; data collection protocols, guidelines, and protocols to pose data-informed questions and design evidence-informed interventions. 

Among the three essential components mentioned in Step 3, we first determined the initial composition of the committee of stakeholders, in consultation with the COI. Then, we recruited the project manager, data scientist, and sociologist. The project manager and data scientist engaged with the stakeholders to understand the stakeholders' data. The city made individual data-sharing agreements with the stakeholders, and the project manager collected their data. These datasets were later visualized by the data scientist on \textit{Sbocciamo Torino} data dashboard. The data scientist built an initial dashboard prototype, which was refined in response to stakeholder feedback. These visualizations were not comprehensive enough to offer a complete picture of the phenomenon of youth deviance in the city, but they got the ground moving. Along with the dashboard, we developed the procedures for the meeting of the committee and ended the development phase.

\subsection*{Step 6: Training of Users}\label{sec:user-training}

The members of the committee of stakeholders were trained in detail about the functioning of the civic tool and introduced to the dashboard and collaborative data analysis methodologies.

Training consists of two parts. The first part focused on familiarizing participants with the structure and operations of the stakeholder committee, including the use of the data dashboard. A simulated committee meeting helped them practice collaborative data analysis and effective in-meeting data communication. Training materials, prepared in Italian, included interactive tools, customized dashboards, and structured prompts to guide the analysis and presentations. The second part covered gender-inclusivity training along with ethical, social, and legal responsibilities in line with national and international gender equality standards. Participants actively engaged in discussing strategies for integrating inclusivity into \textit{Sbocciamo Torino}, supported by practical examples.

\section{Final Design of \textit{Sbocciamo Torino} (RQ1)}\label{sec:components}

In this section, we describe \textit{Sbocciamo Torino} components developed in Step 5. The primary goal of \textit{Sbocciamo Torino} was to understand and co-design intervention proposals aimed at preventing youth deviance in Turin. \textit{Sbocciamo Torino} consists of three main components: \textit{Sbocciamo Torino} committee, a digital data dashboard, and regular co-designing meetings.
Figure~\ref{fig:ST-functioning} summarizes how the components interact.
% The functioning and interrelationship between the components of \textit{Sbocciamo Torino} is summarized by Figure~\ref{fig:ST-functioning}.
The multi-stakeholder committee convenes periodically to discuss and draft intervention proposals by collaboratively analyzing data through a dashboard that visualizes information related to youth deviance. Proposals are forwarded to municipal councilors and discussed in extended committee meetings, which include the councilors or members of their staff. These meetings will facilitate planning and implementation throughout the city.

\subsection{\textbf{\textit{Sbocciamo Torino} Committee}}

The multi-stakeholder committee is composed of a diverse group of actors representing various sectors such as law enforcement, social services, education, and NGOs. The committee consists of both permanent and invited members, where some organizations attend every meeting, while others are invited to the meetings as the need arises. The permanent members include local police, local health, social, and education services, the prisoners’ ombudsperson, and NGOs working with the youth.
This diverse composition brings together different expertise and provides a holistic understanding of the issue of youth deviance in the city. The collective input of the members helps in creating well-rounded and culturally sensitive interventions. 
The organizations of the committee members included:

\begin{enumerate}
    \item Local police (from various units);
    \item Social services department of the city;
    \item Health promotion office of the city;
    \item Local health services; 
    \item Office of the prisoners’ ombudsperson;
    \item Regional anti-doping center;
    \item NGO advocating for children's and minors' rights;
    \item NGO promoting sports and cultural activities;
    \item NGO supporting individuals facing social marginalization, addiction, and isolation. 
\end{enumerate}

The committee is assisted by three supporting members, who have specific roles but do not actively take part in co-designing intervention proposals. They are (1) the \textit{project manager} to coordinate the implementation and facilitate the meetings; (2) the \textit{data scientist} to collect, create, and support the data, visualization, and analysis components of the tool; and (3) the \textit{sociologist} who provides a sociocultural perspective on youth deviance, interpreting data within the broader societal context.

\subsection{\textbf{\textit{Sbocciamo Torino} Dashboard}}\label{sec:dashboard}

The second component of \textit{Sbocciamo Torino} is a digital data dashboard. It visualizes data related to youth deviance in the city and serves as a support system for evidence-based decision-making.
% \textit{Sbocciamo Torino} does not collect new data directly; it uses anonymized data provided by the committee members. The data visualized on the dashboard includes both the stakeholders’ data and the Open Data obtained from the city of Turin:
The dashboard uses anonymized data supplied by committee members and the City Open Data:

\begin{itemize}
    \item \textbf{Open Data} of Turin on demographics (neighborhood population separated based on age, sex, civil status, nationality and support from the city, public spaces (pedestrian areas and green areas), and services (markets, cinemas, museums, libraries, sports facilities, police stations) provided by the city.

    \item \textbf{Stakeholder data} voluntarily provided by committee members.
\end{itemize}

Table~\ref{tab:st-dashboard-data} contains the data provided by\textit{Sbocciamo Torino} committee, which is limited to the first implementation of the civic tool and is not a comprehensive list of datasets. 

%TC:ignore
\begin{table}[ht]
  \caption{Data volunteered by the Initial Committee Members of \textit{Sbocciamo Torino}} \label{tab:st-dashboard-data}
  \begin{tabular}{p{0.02\linewidth}p{0.3\linewidth}p{0.55\linewidth}}
    \toprule
    \textbf{No.} & \textbf{Stakeholder} & \textbf{Data Description} \\ \midrule

    1 & Educational Services Division 
    & Anonymized data of non-attendance or non-compliance in schools. \\
    
    2 & Local Police 
    & Anonymous survey data on experiences of youth aggression experienced by the residents. The survey contains data on frequency, age, and gender of the aggressor(s), location and theft, and the respondent's gender, age at the time of the incident, and present age. \\
    
    3 & Office of the prisoners’ ombudsperson
    & Anonymized data about young detainees. It is a summary sheet providing the number of juveniles based on birth year, judgment status, crime committed, country of origin, city of residence, qualification, activities before prison, age group, previous sentences, and city of crime. \\
    
    4 & Social Services Division
    & The socio-economic and demographic data is obtained through open data, and the social services office will not provide any additional data. \\

    5 & NGO promoting sports and cultural activities 
    & Data on members of the NGO, with anonymized demographics and sports/activities practised. \\

    6 & NGO supporting individuals facing social marginalization, addiction, and isolation
    & Anonymized data about people receiving assistance from the NGO, including demographics, education levels, and current status of assistance.  \\

  \bottomrule
\end{tabular}
\end{table}
%TC:endignore

The city already owned and locally hosted a digital platform that allowed for the creation and maintenance of dashboards. This platform was presented to stakeholders during the concept validation workshop, and they confirmed its suitability for the task. So, given that the cost of using this platform was minimal, we chose that it could provide good support for \textit{Sbocciamo Torino} dashboard. 
The city's digital platform is SQL-based, and we created \textit{Sbocciamo Torino} database (based on PostgreSQL) to store the tables with the relevant stakeholders and open data. 
% This database is stored locally in the city's servers. 

\begin{figure}[ht]
  \centering
  \begin{subfigure}{0.48\textwidth}
      
\includegraphics[width=1\linewidth]{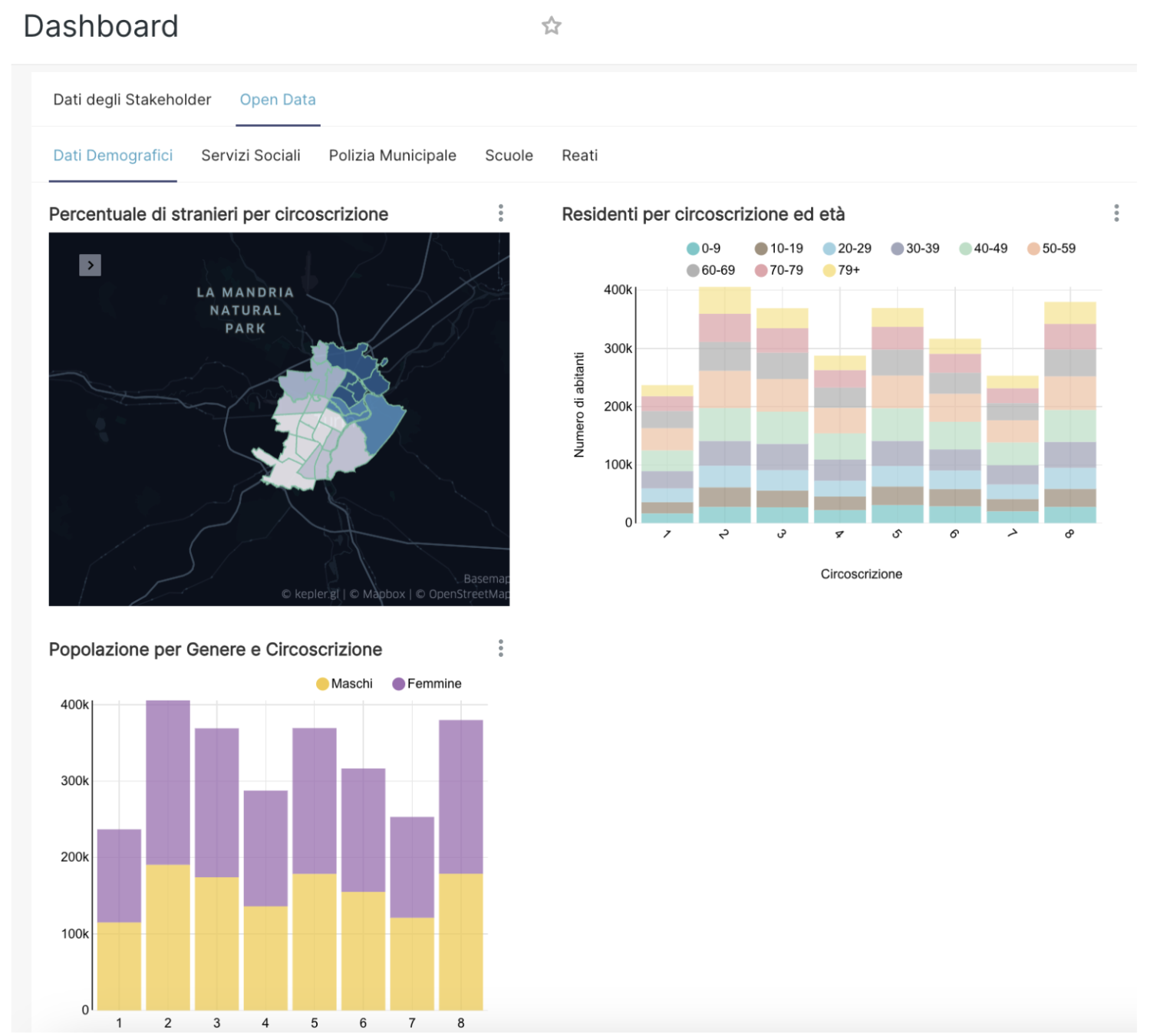}
  \caption{Snapshot of the dashboard on Demographic Data}\label{fig:dashboard-1}
  \Description{Snapshot of \textit{Sbocciamo Torino} Dashboard. This dashboard provides demographic and social data for different city districts. The visualizations include: a map showing the percentage of foreign residents per district, a bar chart of residents segmented by age and district, and a gender distribution by district.}
  \label{fig:ux}
  \end{subfigure}
\hfill
  \begin{subfigure}{0.48\textwidth}
      \includegraphics[width=1\linewidth]{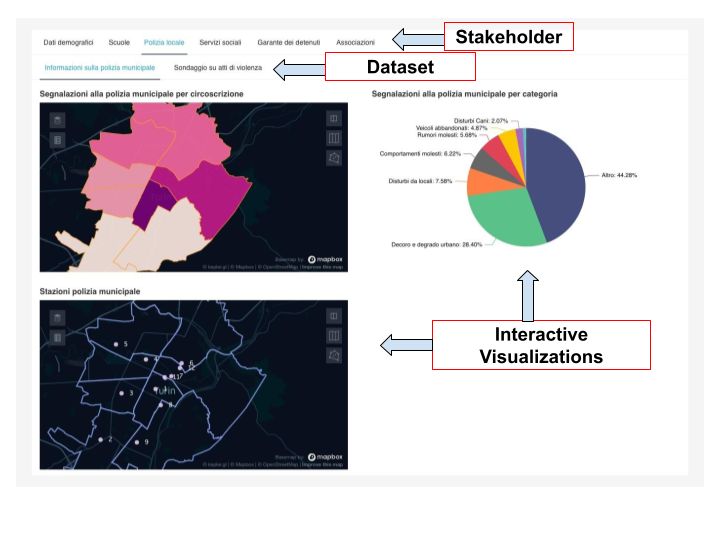}
  \caption{Snapshot of the dashboard on Local Police Data}\label{fig:dashboard-2}
   \Description{Snapshot of \textit{Sbocciamo Torino} Dashboard. This dashboard provides a visualization of local police data for different city districts. The visualizations include: a map showing the percentage of police stations per district, a map showing the locations of the police stations, and a pie chart different types of local police units in the city.}
  \end{subfigure}

  \caption{\textbf{Snapshot of \textit{Sbocciamo Torino} Dashboard.}}
  \label{fig:st-dashboard}
\end{figure}

The dashboard was then built on the existing digital data platform. The dashboard has a modular and interactive design at different levels: Source of data (open or stakeholder) ---> Stakeholder ---> Dataset (Figure~\ref{fig:dashboard-2}).  In particular, the dashboard allows for several types of data visualizations, including histograms, bar plots, pie charts, map-based charts, time series, and word clouds, among others.
Figure~\ref{fig:st-dashboard} represents two snapshots of the \textit{Sbocciamo Torino} dashboard.

%TC:ignore
%TC:endignore

\subsection{\textbf{\textit{Sbocciamo Torino} Meetings}} \label{sec:committee-meeting}

%TC:ignore
\begin{figure}[ht]
  \centering
\includegraphics[width=0.7\linewidth]{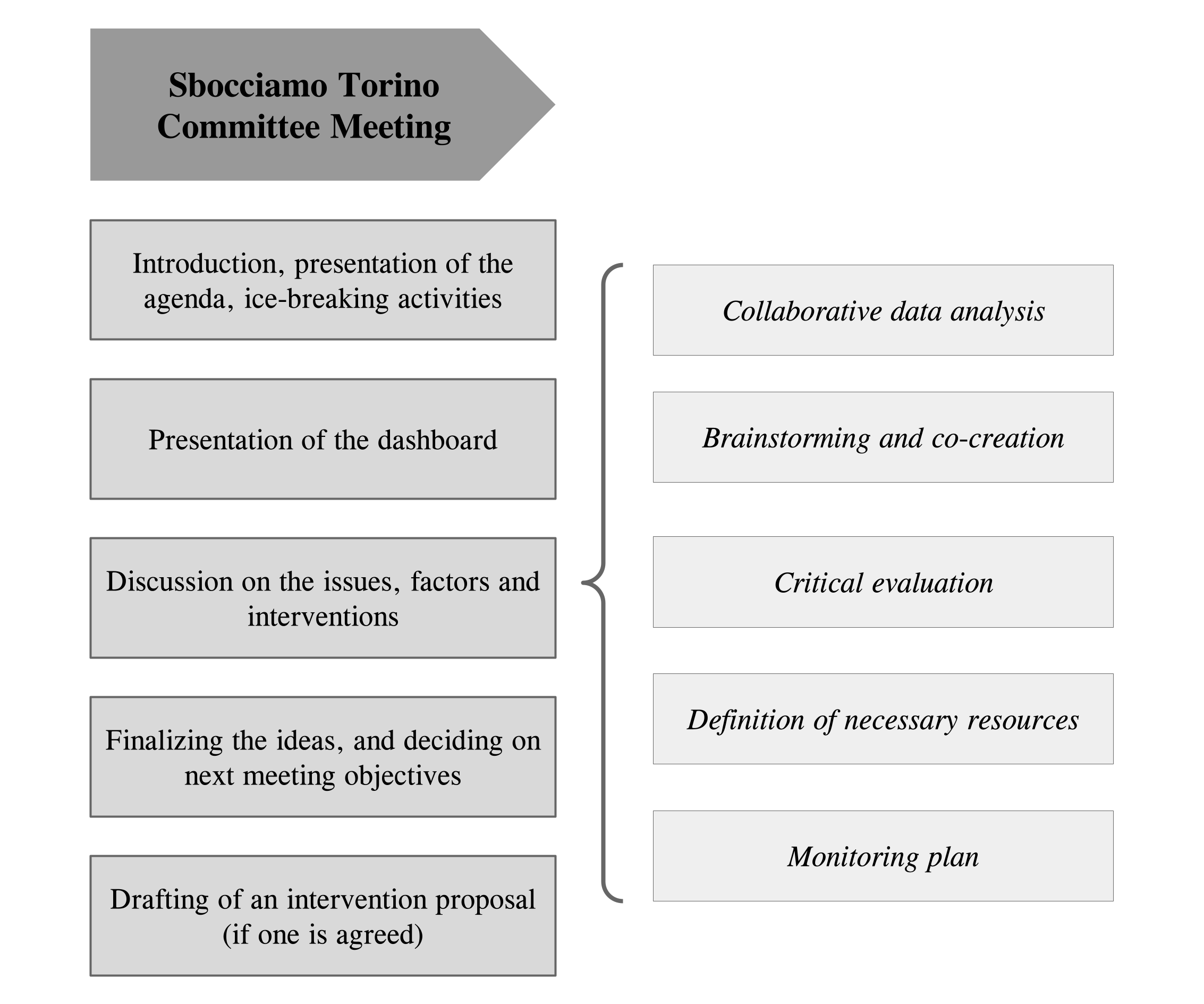}
  \caption{Plan of a \textit{Sbocciamo Torino} committee meeting}\label{fig:ST-meeting}
  \Description{Structure of \textit{Sbocciamo Torino} Committee Meeting. This flowchart presents the typical agenda and activities of an \textit{Sbocciamo Torino} committee meeting. It begins with an introduction and agenda presentation, followed by a dashboard overview, discussion on relevant issues, and finalization of ideas and objectives. The process includes collaborative data analysis, brainstorming, critical evaluation, resource definition, and the development of a monitoring plan. If consensus is reached, an intervention proposal may be drafted for further action.}
\end{figure}
%TC:endignore

% The meetings of the committee of stakeholders are organized by the project manager, occurring at least once every six months, or at the request of the majority of members. Additional participants are brought in based on the meeting agenda or suggestions from stakeholders.

Each \textit{Sbocciamo Torino} committee meeting (\textasciitilde{}100 minutes) is structured for co-designing interventions.
% The meetings of the \textit{Sbocciamo Torino} committee, which take approximately 100 minutes, are composed of different parts to allow for the co-design of intervention proposals.
The structure of each meeting is shown in Figure~\ref{fig:ST-meeting} with a timed agenda in Appendix~\ref{sec:appendix-committee-meeting-process}. The meeting is based on both the collaborative analysis of the data visualized in the dashboard and the collective hands-on experience of the stakeholders. The participants discuss potential interventions to address the issues on the agenda, and, if they come to a consensus, they outline a proposal for the intervention. A two-thirds majority of the committee members is necessary to approve a decision.
% A detailed procedure for a committee meeting, including a precise time allocation for each activity is provided in Appendix~\ref{sec:appendix-committee-meeting-process}. 
Once an intervention proposal is agreed upon, the project manager drafts the proposal and shares it with the offices of the municipal councilors.

\subsection{\textbf{\textit{Sbocciamo Torino} Extended Committee Meetings}} \label{sec:extended-committee-meeting}

%TC:ignore
\begin{figure}[ht]
  \centering
\includegraphics[width=0.85\linewidth]{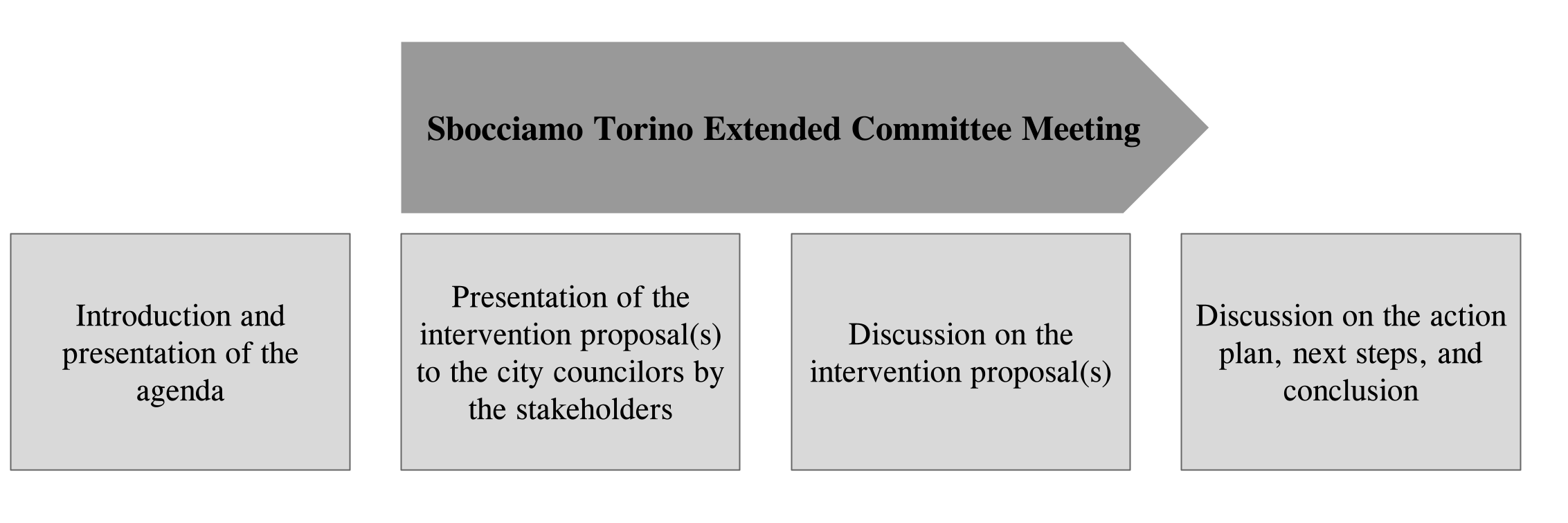}
  \caption{Plan of an \textit{Sbocciamo Torino} extended committee meeting}\label{fig:ST-extended-meeting}
  \Description{Structure of \textit{Sbocciamo Torino} Extended Committee Meeting. This diagram outlines the flow of an extended committee meeting for \textit{Sbocciamo Torino}. It begins with the introduction and agenda presentation, followed by stakeholders presenting intervention proposals to the city councilors. The meeting proceeds with discussions on the proposals and concludes with a review of the action plan, next steps, and final remarks. This structured approach facilitates effective communication and decision-making among stakeholders and city officials.}
\end{figure}
%TC:endignore

After an intervention proposal is submitted to the municipal councilors' office, the \textit{Sbocciamo Torino} committee meets them.
% project manager organizes a meeting of the extended committee, which includes both \textit{Sbocciamo Torino} committee members and the municipal councilors.
The committee presents the intervention proposal and discusses its feasibility and the next steps for the implementation. Figure~\ref{fig:ST-extended-meeting} shows the unfolding of a one-hour meeting of the extended committee, and the process is in Appendix~\ref{sec:appendix-extended-committee-meeting-process}.
After the meeting, the project manager prepares and shares the meeting minutes with all extended committee members. In addition, when the action plan is approved, the appropriate stakeholders should plan the implementation strategy and further action.

\section{Results: Implementation and Design Evaluation of \textit{Sbocciamo Torino} (RQ2)} \label{sec:results}

Given the participatory and iterative nature of our work, we present the results in two parts. First, we describe the real-world demonstration of \textit{Sbocciamo Torino} at the inaugural committee meeting. Second, we present findings from two evaluations on usability, stakeholder perceptions, and organizational conditions. We used a survey and an open-ended questionnaire to collect insights into both the implementation and reception of the civic tool. 
% These results will help us understand the practical considerations, challenges, and opportunities that shaped the participatory design and early use of the tool in context, thus addressing our second research question (RQ2).
Together, they address RQ2 by revealing practical considerations, challenges, and opportunities.

\subsection{Step 7: Demonstration}\label{sec:Demonstration}

This subsection presents the outcomes of \textit{Sbocciamo Torino} committee’s inaugural meeting that marked the tool's real-world implementation.
It was organized and moderated by the project manager, who brought together 12 committee members, the data scientist, and the sociologist. As this was the inaugural meeting, representatives from the municipal councilors for security and welfare also attended as observers.
Prior research~\cite{Crawford2023} and stakeholder input surfaced that addiction prevention plays a crucial role in addressing youth deviance. Programs in other cities often engage schools, families, and communities through multifaceted interventions. Reflecting this, the agenda of the first committee meeting focused on “understanding the intersection of youth deviance and drug abuse.”

The meeting opened with a welcome from the municipal councilor for security, followed by a brief presentation from the project manager outlining the agenda and workflow. 
% The data scientist then introduced \textit{Sbocciamo Torino} dashboard, which, in its initial version, provided an overview of currently available data to help identify gaps and guide the committee’s future priorities.
The data scientist presented the initial dashboard to highlight current data and gaps.

The core of the meeting consisted of brainstorming and discussion about the types of data still needed and the specific issues the committee should address under this agenda. Members raised several critical points, including the misuse of psychotropic drugs among youth --- particularly among those lacking self-regulation skills, as noted by the representative of the prisoners’ ombudsperson's office --- and the rise in drug-related challenges since the COVID-19 lockdowns, as highlighted by the local police. Participants also stressed the importance of longitudinal data collection to observe individuals during and after incarceration, while addressing challenges related to anonymity and data continuity.
An NGO representative proposed focusing on psychotropic drug abuse both inside and outside of prison settings, emphasizing the need to address multi-substance use (alcohol, tobacco, psychotropics) and to involve street-based and field-level interventions. his proposal was based on an existing study conducted by the prisoners' ombudsperson's office regarding drug abuse in prisons, and qualitative observations made by NGO representatives and police officers, as well as the representative of the regional anti-doping center. The committee reached a consensus on the centrality of this issue to the local context of youth deviance.

The meeting concluded with an agreement to prioritize addictions --- particularly psychotropic drug abuse --- in upcoming committee activities. The group called for a feasibility analysis to improve data collection and expand the dashboard’s integration with quantitative data related to addiction and drug use. With a clearly defined focus and improved data input, future meetings will move toward drafting intervention proposals, future meetings will draft intervention proposals for the municipal councilors. This demonstration served as a practical test of how the tool facilitated collaborative agenda-setting, data exploration, and issue prioritization.

\subsection{Step 8: Design Evaluation} \label{sec:evaluation}

This subsection presents evaluation results that reflect stakeholder experiences at two key stages of implementation: post-training and post-demonstration, after the training session and following the first demonstration. Using both quantitative and qualitative methods, these evaluations assessed the usability of the tool, stakeholder understanding of their roles, and perceived organizational and infrastructural challenges. The results shown here capture both structured responses (via Likert-scale survey) and open-ended reflections, which provide perspectives on the strengths and limitations of the civic tool’s participatory design and integration.

\subsubsection{Design Evaluation 1: Post Training} \label{sec:result-post-train}

Here we present the post-training survey results. Nineteen participants (Table~\ref{tab:part1-eval-users}) answered Likert-scale questions (Appendix~\ref{sec:appendix-survey-january}) on their understanding of the tool, clarity regarding their roles, and perceptions of the participatory and multilateral approach. 

%TC:ignore
\begin{table}[ht]
  \caption{Details of the participants who took part in the post-training evaluation.}
  \label{tab:part1-eval-users}
  \begin{tabular}{ccc}
    \toprule
    \textbf{ID} & \textbf{Type of Organization} & \textbf{Number of Participants} \\
    \midrule
     1 & Third Sector (NGO, association, cooperative) & 4 \\ \\

     2 & Municipal Offices & 4 \\ \\

     3 & Local Police (from various units) & 11 \\

  \bottomrule
\end{tabular}
\end{table}
%TC:endignore

\begin{figure}[ht]
  \centering
\includegraphics[width=0.8\textwidth]{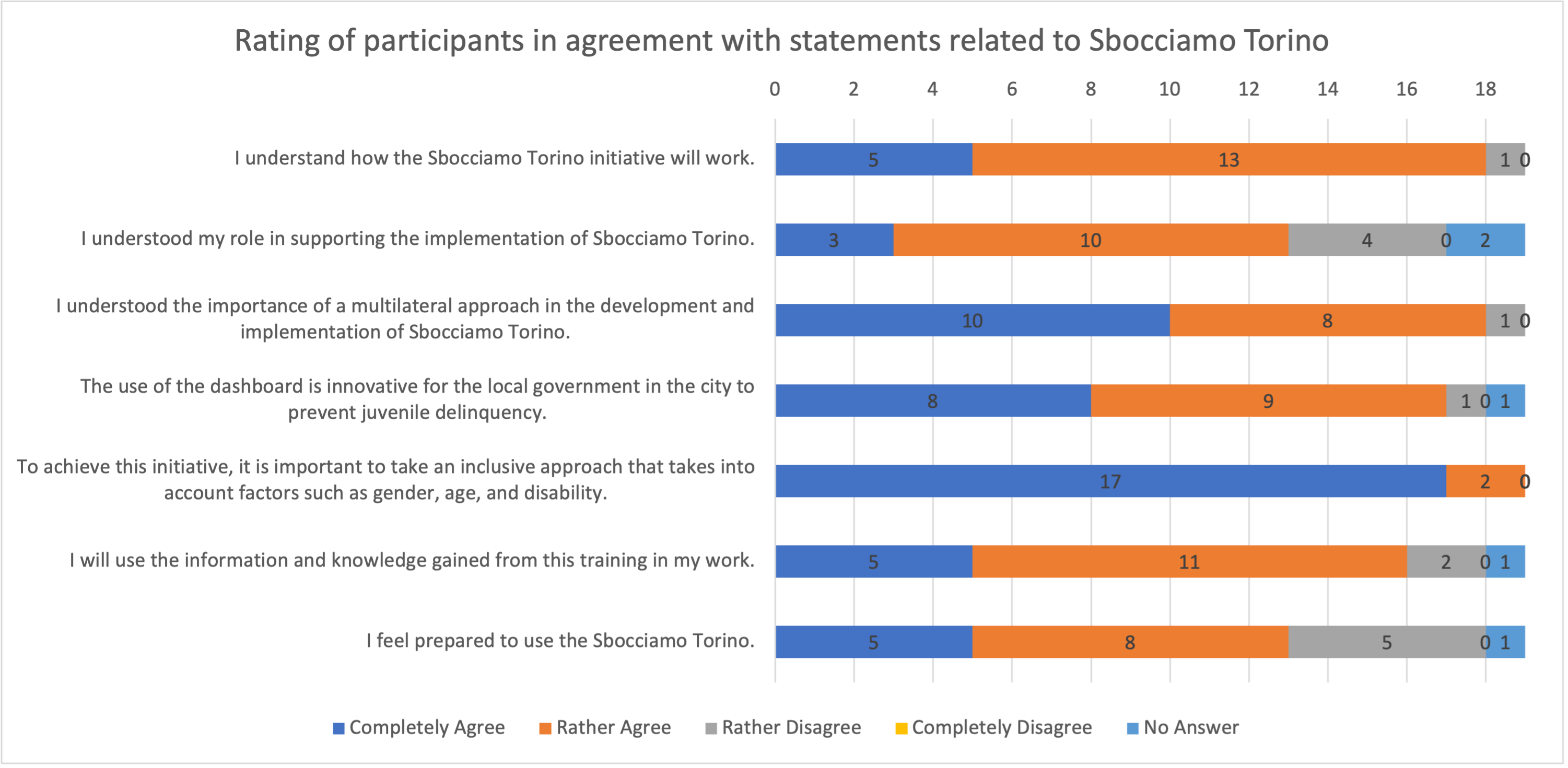}
  \caption{Rating of participants in agreement with statements related to \textit{Sbocciamo Torino}}\label{fig:part-1-eval}
  \Description{Participant Ratings on \textit{Sbocciamo Torino}. This bar chart illustrates participants' agreement levels with statements regarding \textit{Sbocciamo Torino}. It covers their understanding of the tool's functionality, their roles, and the importance of a multilateral and inclusive approach in its development. Additional statements reflect the participants' views on the tool's innovative use of a dashboard for local governance and its relevance for preventing juvenile delinquency. Agreement levels range from 'Completely Agree' to 'Completely Disagree,' highlighting overall support for the tool's implementation and participants’ preparedness to utilize it.}
\end{figure}

The results (Figure~\ref{fig:part-1-eval}) show generally positive attitudes toward the tool and its design process, while also revealing areas where additional support and clarification were needed --- especially around role definition and confidence in dashboard usage. Specifically, 18 respondents understood how the tool worked, though only 13 felt fully confident about their specific roles. The reason for low confidence among a few members was because of their absence from the training.
All participants valued the inclusive approach throughout solution development. Further, 18 participants agreed that a multilateral approach is crucial for the success of \textit{Sbocciamo Torino}, and one participant expected a greater push from the city. These responses highlight the participants' appreciation for the inclusive and multilateral strategies employed by \textit{Sbocciamo Torino}. 17 out of 19 participants viewed using the data-assisted multi-stakeholder co-design approach to support urban youth at risk by assisting local authorities as innovative. Sixteen participants agreed that the knowledge gained from their involvement in the process would be valuable in their work. These responses indicate that stakeholders and participants recognized and valued the expertise and innovation brought by their experience with \textit{Sbocciamo Torino}.

%TC:ignore
%TC:endignore

Overall, the survey findings indicate that stakeholders recognized the importance of inclusivity and collaboration, which were the civic tool's primary objectives. 
However, the findings also reveal that a few participants require a better understanding of their roles and how to effectively use the dashboard.

\subsubsection{Design Evaluation 2: Post Demonstration} \label{sec:result-post-demon}

\begin{figure}[ht]
  \centering
\includegraphics[width=0.8\linewidth]{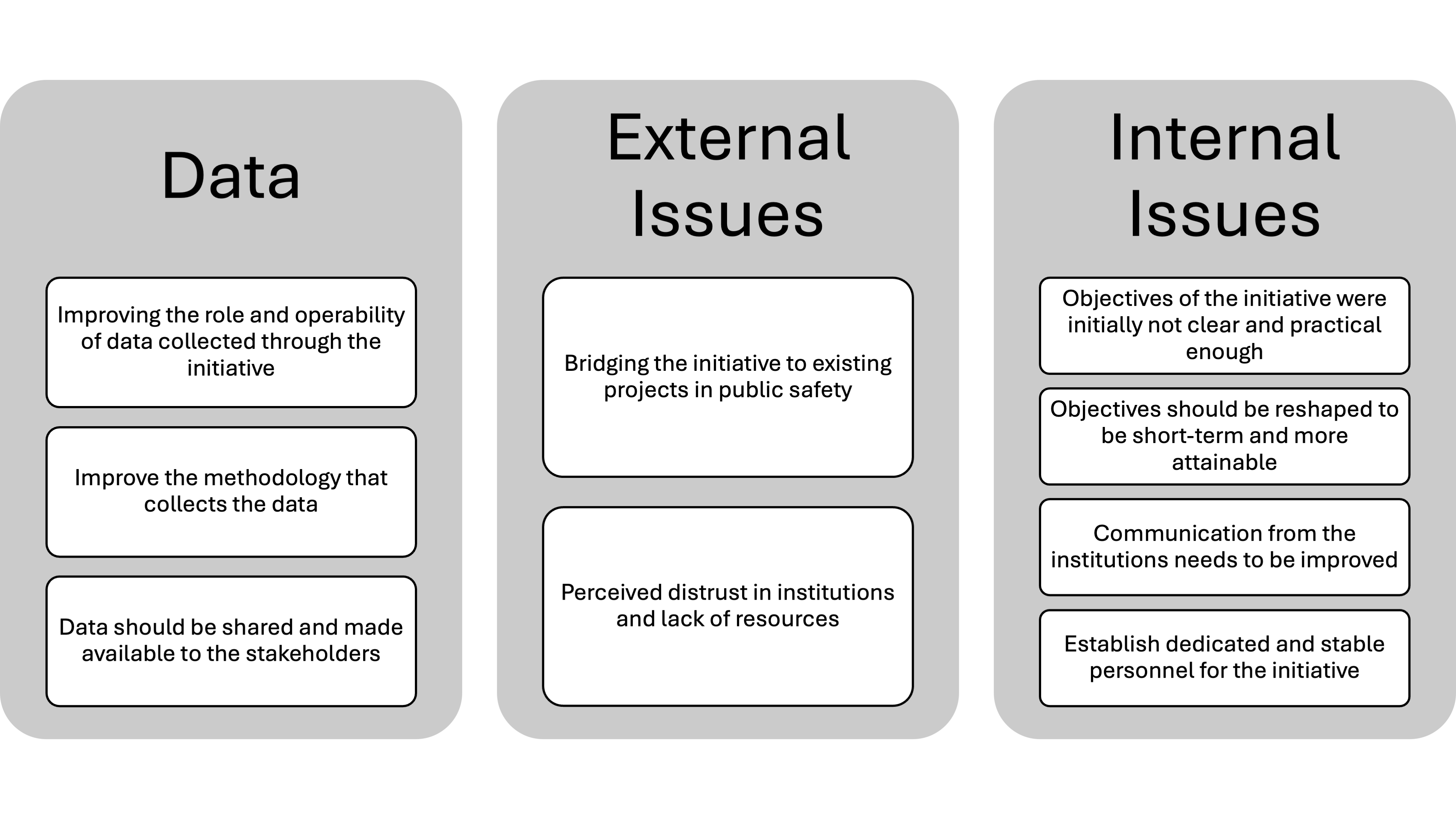}
  \caption{Code tree of answers to open-ended questionnaire of the second component of the evaluation.
}\label{fig:code-tree}
  \Description{Open Coding Analysis of \textit{Sbocciamo Torino} evaluation. This figure presents an open coding analysis of key challenges categorized into three main areas: Data, External Issues, and Internal Issues. The Data section highlights the need to improve data collection methods, operability, and sharing among stakeholders. External Issues focus on bridging the civic tool with existing public safety projects and addressing institutional distrust and resource limitations. Internal Issues emphasize the necessity for clearer, short-term objectives, better communication from institutions, and the establishment of stable, dedicated personnel to support the civic tool's success.}
\end{figure}

The second part of the evaluation was a seven-question open-ended questionnaire after the first meeting of the stakeholder committee (Step 7). It gathered qualitative insights into the clarity of the goals, challenges faced during development, limitations, and suggestions for improving participation, communication, and future implementations.
Respondents included the project manager, data scientist, sociologist, representatives from the municipal councilors for security and welfare, and an external observer (Table~\ref{tab:part2-eval-users}). The responses were analyzed using open coding and grouped into three thematic categories: \textit{data-related insights, internal organizational issues, and external institutional factors} as shown in the code tree in Figure~\ref{fig:code-tree}. The results reveal both enthusiasm for the tool’s collaborative potential and concerns related to communication, goal-setting, and long-term planning.

%TC:ignore
\begin{table}[ht]
  \caption{Participants who took part in \textit{Sbocciamo Torino}'s demonstration evaluation.}
  \label{tab:part2-eval-users}
  \begin{tabular}{p{0.015\linewidth}p{0.5\linewidth}p{0.25\linewidth}p{0.12\linewidth}}
    \toprule
    \textbf{ID} & \textbf{Organization} & \textbf{Role} & \textbf{Number of Participants} \\
    \midrule
     1 & Regional anti-doping center & Stakeholders' committee & 1 \\

     2 & Office of the prisoners’ ombudsperson & Stakeholders' committee & 1 \\

     3 & Local health services and Health promotion office of the city & Stakeholders' committee & 3 \\

     4 & NGO advocating for children's and minors' rights & Stakeholders' committee & 2 \\

     5 & NGO supporting individuals facing social marginalization, addiction, and isolation & Stakeholders' committee & 2 \\

     6 & Local Police (from various units) & Stakeholders' committee and Project Manager & 4 \\
    
     7 & Research institute & Data Scientist & 1 \\

     8 & University & Sociologist & 1 \\

     9 & Office of the municipal councilor for security & Observer & 1 \\

     10 & Office of the municipal councilor for welfare & Observer & 1 \\

     11 & National NGO for urban security & Observer & 1 \\

  \bottomrule
\end{tabular}
\end{table}
%TC:endignore

In the first category, \textit{Data}, many stakeholders focused on the role of data in the civic tool, particularly how data sharing could enhance decision-making processes and facilitate it. The second category, \textit{External Issues}, encompassed issues and solutions identified by stakeholders that were external to the design itself. The third category addressed \textit{internal challenges }stakeholders faced during their involvement with the design of the civic tool.

All respondents indicated that they understood the objectives of the civic tool, with some crediting the demonstration session for this clarity. For example, the representative of the local health services (\textit{ID 3, Table~\ref{tab:part2-eval-users}}) stated that the scope of the tool was clear ``\textit{especially in its goal of data sharing across different organizations}'', and the representative of an NGO (\textit{ID 5, Table~\ref{tab:part2-eval-users}}) mentioned that ``\textit{all the goals were explained in a sufficiently clear way.}'' 

It was noted that there was an initial lack of interest in the participatory process. This was due to the organizational issues at the city level in the scheduling of the workshops and training sessions. While the reasons are outside the scope of the civic tool, the evaluation shows the need for clearer communication of the objectives and better coordination between the institutions involved. Respondents recommended improving communication strategies to better disseminate information and using the tool's results to showcase its benefits to attract more stakeholders and participants. The representative of the office of city councilor for welfare (\textit{ID 10, Table~\ref{tab:part2-eval-users}}) stated that ``\textit{associations would feel more engaged if they were facing clear goals with respect to their role},'' and the representative of an NGO (\textit{ID 4, Table~\ref{tab:part2-eval-users}}) stated that it would be beneficial to ``\textit{have short-term goals that are easy to reach so that it is possible to show in a short time the effectiveness of the tool}.'' This feedback contrasts with the challenge of ``implementing continuous, structured programs rather than short-term projects'' (Step 2, Section~\ref{sec:dt-workshop}), indicating that while some participants value long-term impact, others see short-term results as crucial. Although it is not the feedback from the same person, the discrepancy reflects varying organizational priorities or perspectives. It reinforces the need for adaptable communication and goal-setting strategies across stakeholder groups.

They also recommended enhancing the usability of the collected data, often advocating for the establishment of dedicated personnel to develop sustainable, long-term planning for the civic tool. One participant said ``\textit{[This civic tool] will improve the transversality of the interventions and help in creating joint programs with the associations}'' (\textit{ID 10, Table~\ref{tab:part2-eval-users}}). At the same time, a representative of the local health services (\textit{ID 3, Table~\ref{tab:part2-eval-users}}) noted that ``\textit{[There is] lack of resources needed for synthesis (analysis)}.'' Respondents valued this collaborative co-design approach as it could promote interest among potential stakeholders and build lasting relationships. A representative of the local police (\textit{ID 6, Table~\ref{tab:part2-eval-users}}) valued an ``\textit{enhanced collaboration across sectors and departments within the municipality}.''

Overall, respondents highlighted the need for improved institutional communication and augmented support for data analysis. Improving these aspects could boost more support and facilitate the integration of \textit{Sbocciamo Torino} with existing local and national projects.

%TC:ignore
%TC:endignore

\section{Discussion} \label{sec:discussion}

This study contributes to ongoing research on participatory civic technologies by detailing the development of \textit{Sbocciamo Torino} --- a collaborative decision-making system grounded in participatory design and Research through Design (RtD) principles. Positioned within broader discussions around socio-technical systems and community-centered innovation, this work illustrates the potential and complexities of co-developing civic tools for collaborative data-driven approaches.

Through the design of \textit{Sbocciamo Torino}, we show the use of methods, such as user-centered design, iterative testing, and participatory approaches, in developing civic tools spanning physical and digital spaces to bring together actors with various areas of expertise~\cite{Luther2009, Kriplean2012, Haesler2021, Brandon2021opencivic}. We underline the need for participatory process and iterative testing to surface the problems at early stages as discussed in previous works about the launching of the collaborative efforts, hierarchy of public administrations, and the formalities of stakeholder interaction and ambiguity among stakeholder representatives regarding their expected function~\cite{Merkel2004, Saxena2020, Bratteteig2016, Harrington2019, Bryson2015, Leonardi2023}. In our case, we identified literacy gaps in data, technology, and collaborative analysis, necessary to be addressed before full-scale implementation can take place, and provided alternative measures to balance them. However, the evaluation of the demonstration also showed that without sustained engagement, any civic tools risk becoming irrelevant or underutilized~\cite{Bratteteig2016-1, McGuinness21, Gilman2017}. We contribute to the ongoing work on digital civics, multi-stakeholder cooperation with varying priorities and expertise, and group decision-making~\cite{Boehner2016, Lindley2017, Choi2017, Stoll2012}. Specifically, we provide a case study on the effective use of participatory design in the development of civic tools and their use to support collaborative decision-making. 

\subsubsection*{\textbf{Reflections on the Participatory Processes}}

The development of \textit{Sbocciamo Torino} illustrates the complexities of participatory design in real-world civic settings. While participatory methods are often associated with inclusion, transparency, and shared ownership, our experience highlights the importance of enabling sustained engagement through organizational commitment, alignment with institutional priorities, and the creation of trusted spaces for exchange and dialogue.

One of the key insights from our process was the importance of procedural integration. From the early Design Thinking workshop (Step 2, Section~\ref{sec:dt-workshop}), it became evident that stakeholders valued visible access to decision-making structures. This was reinforced in later stages of the process (e.g., Step 4), where the inclusion of city councilors and the formalization of committee processes were seen as essential for legitimacy. This echoes findings that emphasize the importance of making participatory infrastructures legible and actionable to institutional actors~\cite{ledantec2012, Crivellaro2019, Bratteteig2016}.

At the outset of the project, we articulated a set of informal aspirations for the civic tool together with the stakeholders. These included enabling collaboration across organizational boundaries, supporting structured data sharing, and embedding community deliberation within municipal processes. While these goals provided orientation, they were not used as fixed milestones or predefined outcomes. As the work progressed, it became clear that what mattered most were the relationships built, the procedural structures co-developed, and how the tool began to embed itself into institutional workings. This resonates with prior work on infrastructure in HCI, which suggests that participatory technologies evolve through ongoing negotiation and alignment with local practices and constraints~\cite{Vlachokyriakos2016, Ledantec2013}.

A central challenge that shaped the process was the disparity in data capabilities among stakeholders. While some organizations had dashboards and structured data pipelines, others relied on spreadsheets or had no formal data practices. This unevenness reflects a broader tension in civic technology work, where technological capacity and organizational readiness vary widely~\cite{Boehner2016, Gilman2017}. To address this, the design included support roles (a data scientist and a project manager) who provided hands-on guidance in preparing and sharing data. This practical scaffolding helped establish a baseline for collaboration without excluding less digitally mature organizations --- an approach also discussed in the context of inclusive civic platforms~\cite{Mahyar2017, Puussaar2018}.

The role of the city was both enabling and delimiting. Institutional support lent credibility to the civic tool and enabled resource allocation (e.g., personnel, hosting infrastructure), but it also shaped the boundaries of participation. The city determined which stakeholders could be formally involved, what types of data could be shared, and how recommendations could be communicated. Balancing institutional constraints with the goals of inclusive participation remained a persistent tension, a well-documented challenge in participatory governance and urban civic design~\cite{Crivellaro2019, Stoll2012, Lindley2017}.

Hence, the participatory process behind \textit{Sbocciamo Torino} underscores that civic co-design is as much about maintaining relational and organizational continuity as it is about generating new systems or interfaces, as it requires attending to institutional frictions and stakeholder asymmetries throughout the entire lifecycle of the project.

\subsubsection*{\textbf{Stakeholder Engagement}}

The design evaluation revealed that the solution was well received by the stakeholders, and it became apparent that the process of involving stakeholders in the development was successful. In addition to the goal of evidence-based decision-making, stakeholders recognized the resulting benefits and synergies, such as the definition of common goals and the strengthening of mutual trust. However, engaging them throughout the process proved to be hard. The difficulty arises due to limitations in the stakeholders’ operational and communication capabilities. Firstly, the stakeholders involved in the process have different primary jobs, and finding a common time to meet is in itself a big task. While we productively engaged with individual stakeholders on some occasions, it was necessary to meet them all together at certain times during the process. The long and time-consuming nature of the participatory design, combined with personnel changes within the local police, who were acting as liaisons between the researchers and the stakeholders, made it challenging for continuous communication between researchers and stakeholders~\cite{Robertson2012}. At the same time, nurturing personal relations with the stakeholders is paramount in human-centric processes, as it establishes trust among the stakeholders and promotes the creation of a smooth working environment. The changes of personnel also affected this process, as it took additional time to regroup the replacements for each of the corresponding stakeholders and convince them to be part of the process~\cite{Zhang2022, Gottlieb2024}. 

These challenges were evident during the training session (Step 6, Section~\ref{sec:user-training}). Originally scheduled for November 2023, the training had to be postponed to January 2024 due to low participation and operational issues. This delay underscored the need for better coordination and assessment of stakeholders’ data and digital literacy, and collaborative capabilities. The prototyping and training sessions show the need for an extensive exchange and preparation of stakeholders, to gather information on the data that each stakeholder can provide, and to assess their level of literacy on data, digital practices, and group collaboration.

Lastly, if developed in a meaningful way, these forms of participation and collaboration can create a sense of ownership among stakeholders, which can increase commitment on all sides~\cite{Crawford2023}. However, a good implementation of participation formats is crucial: attention must be paid to transparent communication and decision-making. Not taking this into account will likely contribute to frustration and alienation, and will also make cooperation more difficult in future collaborations~\cite{Volkmann2023}. In addition, although \textit{Sbocciamo Torino} benefited from dedicated municipal staff and stakeholder support, the same process can be scaled down for less-resourced cities. Based on our experience, cities can begin with existing open-data portals and spreadsheets for data dashboards and tap into whatever stakeholder networks already exist, regardless of size or technical capacity.

\subsubsection*{\textbf{Trust and Political Support}}

Trust is a key aspect for the efficient participatory design and deployment of \textit{Sbocciamo Torino}~\cite{Oomsels2014}. Establishing trust --- both in institutions and in the tool’s effectiveness --- requires time and consistent political support. As prior research has emphasized, long-term cooperation systems for community organizations must be affordable, transparent, and institutionally supported to remain viable over time~\cite{Haesler2021, Stoll2012}. Institutional transparency, particularly in how committee recommendations are integrated into city decision-making, plays a central role in reinforcing stakeholder trust. The development of a trustworthy relationship is essential to avoid a disconnect between stakeholders and decision-makers. Failing to establish functional inter-personal and inter-organizational trust patterns among stakeholders~\cite{Oomsels2019} risks alienating the latter and undermining the success of the whole participatory process~\cite{Corbett2021}. 

In this context, political support acts as both a foundation and an amplifier of trust. In our case, the political support came from the offices of the municipal councilors for security and welfare.  These councilors publicly endorsed the civic tool and the participatory process, mobilized internal teams to support it, and were consistently present at co-design workshops and public events. They publicly spoke about the benefits brought by the joint work of stakeholders in different internal and external forums --- boosting stakeholders’ confidence in the civic tool.
Ultimately, when paired with meaningful participation and institutional responsiveness, political backing can catalyze the kind of trust necessary for durable collaboration between city officials, NGOs, citizen organizations, and other local actors~\cite{Uslaner2005}.
The experience of \textit{Sbocciamo Torino} not only shows that political support can legitimize participatory processes, but also enables further co-creation and participatory decision-making.

\subsubsection*{\textbf{Limitations}}
Our work has several limitations that should be acknowledged. First, while we aimed for a broad and inclusive participatory process, the stakeholders involved in the design of \textit{Sbocciamo Torino} do not represent the full spectrum of relevant actors. In addition to the committee members, we made efforts to engage other key groups --- including religious organizations, youth committees, journalists, and additional third-sector associations working with young people in the city. However, a combination of practical constraints on the stakeholders' side (as discussed above) and the project’s limited timeline prevented their full involvement.
A particularly important limitation is the absence of young people themselves from the process. Given that the tool is designed to address youth deviance, their perspectives should ideally inform both problem framing and solution design. This omission reflects a known challenge in participatory governance and youth justice research~\cite{Creaney2024}: although policy and design frameworks increasingly recognize the importance of involving youth, legal, ethical, and procedural barriers often prevent direct engagement --- particularly when working within institutional constraints such as those faced by our municipal partners~\cite{Hodson2023, Jardine2012}. Future work should seek to develop ethically appropriate mechanisms to include young people in meaningful, structured, and safe ways.

The designed civic tool does not have the means to function independently. Its current viability relies on the city’s active support: a project manager coordinates operations, the municipality covers operational costs, and the infrastructure is maintained by the city’s digital services. While this institutional backing is essential for integration and long-term adoption, it also creates dependencies that may limit adaptability or continuity if political priorities shift or resources are withdrawn~\cite{Harding2015}. A more sustainable model would allow the tool to secure its own funding and operational autonomy --- possibly through partnerships, grant mechanisms, or civic innovation ecosystems. These concerns are echoed in prior HCI research, which highlights how civic platforms often rely on ephemeral project funding or informal labor, making them vulnerable to institutional change~\cite{Crivellaro2019, ledantec2012, Balestrini2017, Harding2015}. Sustainable participatory infrastructures require not only technical robustness but also stable governance, long-term stewardship, and models for evolving ownership and maintenance over time~\cite{Voida2014}

Lastly, we only conducted the design evaluation in the current work. The current evaluation did not examine the quality and effectiveness of the intervention proposals generated by the committee. Please note that it is only possible to evaluate the committee’s work in the long run, spanning multiple years. Such an evaluation is clearly outside the scope of this paper, but could be the subject of a future publication about this longitudinal aspect.

\section{Conclusion} \label{sec:conclusion}

In this work, we discuss the design, development, and evaluation of \textit{Sbocciamo Torino}, where we adopted a Research through Design approach in collaboration with city administrators, local police units, municipal departments, and NGOs. The civic tool was designed to support structured collaboration around data, enabling stakeholders to deliberate on intervention strategies related to youth deviance --- defined locally as behaviors ranging from minor antisocial conduct to acts of violence and substance abuse. In conclusion, we revisit the two research questions posed at the beginning and provide corresponding answers.

\textbf{\textit{RQ1}: How to co-design participatory civic tools for data-informed and multi-stakeholder collaboration in decision-making to support youth deviance prevention in Turin?} 
We show that participatory civic tools must go beyond co-design workshops or interface feedback: in our case, \textit{Sbocciamo Torino} emerged as a procedural infrastructure, combining a stakeholder committee, a shared data dashboard, and structured deliberative meetings. Its design was directly informed by the need for trust-building across sectors, clarification of institutional roles, and the creation of spaces for joint agenda-setting. Notably, the tool's structure reflects the particular organizational ecology of Turin, in particular the collaboration across loosely connected actors, significant differences in data collection practices, and the need for mediation between civic actors and municipal policy processes.

\textbf{\textit{RQ2:} Which practical considerations, such as institutional processes and stakeholder attitudes, of communities in the city shape the design of such a civic tool?}
We identified several grounded factors that shaped the civic tool’s form and process: stakeholder time constraints, staff turnover (particularly within police units), low data literacy, and the need for institutional legitimacy to ensure long-term engagement. These constraints required us to design for ongoing facilitation, technical support roles (i.e., project manager, sociologist, and data scientist), and gradual capacity-building. Additionally, political endorsement from the city --- while enabling formal uptake --- also constrained the scope of participation, limiting youth engagement due to legal and ethical boundaries.

Our findings contribute to research on communities and technologies by showing how participatory civic tools can be embedded within existing urban governance structures, not just imagined in idealized community contexts. \textit{Sbocciamo Torino} illustrates what it takes to design for real-world inter-organizational collaboration, including the importance of institutional trust, visible feedback loops into decision-making, and adaptive support for stakeholder diversity.
Future work will evaluate the longer-term use and impact of \textit{Sbocciamo Torino}, including how intervention proposals are taken up by city councilors, how youth voices might be included in future iterations, and how the model could be adapted to other domains or cities facing similar urban challenges.

%%
%% The acknowledgments section is defined using the "acks" environment
%% (and NOT an unnumbered section). This ensures the proper
%% identification of the section in the article metadata, and the
%% consistent spelling of the heading.
\begin{acks}
This work was supported by the IcARUS project, funded by the European Union’s Horizon 2020 research and innovation programme under grant agreement No. 882749. We would like to thank all the participants in the city of Turin and our partners at the European Forum for Urban Security (EFUS) for their support.
\end{acks}

%%
%% The next two lines define the bibliography style to be used, and
%% the bibliography file.
\bibliographystyle{ACM-Reference-Format}
\bibliography{bibliography}

%%
%% If your work has an appendix, this is the place to put it.
\appendix
\newpage
\section{Stakeholder Mapping} \label{app:stakeholder-mapping}

%TC:ignore
\begin{figure}[ht]
  \centering
\includegraphics[width=\linewidth]{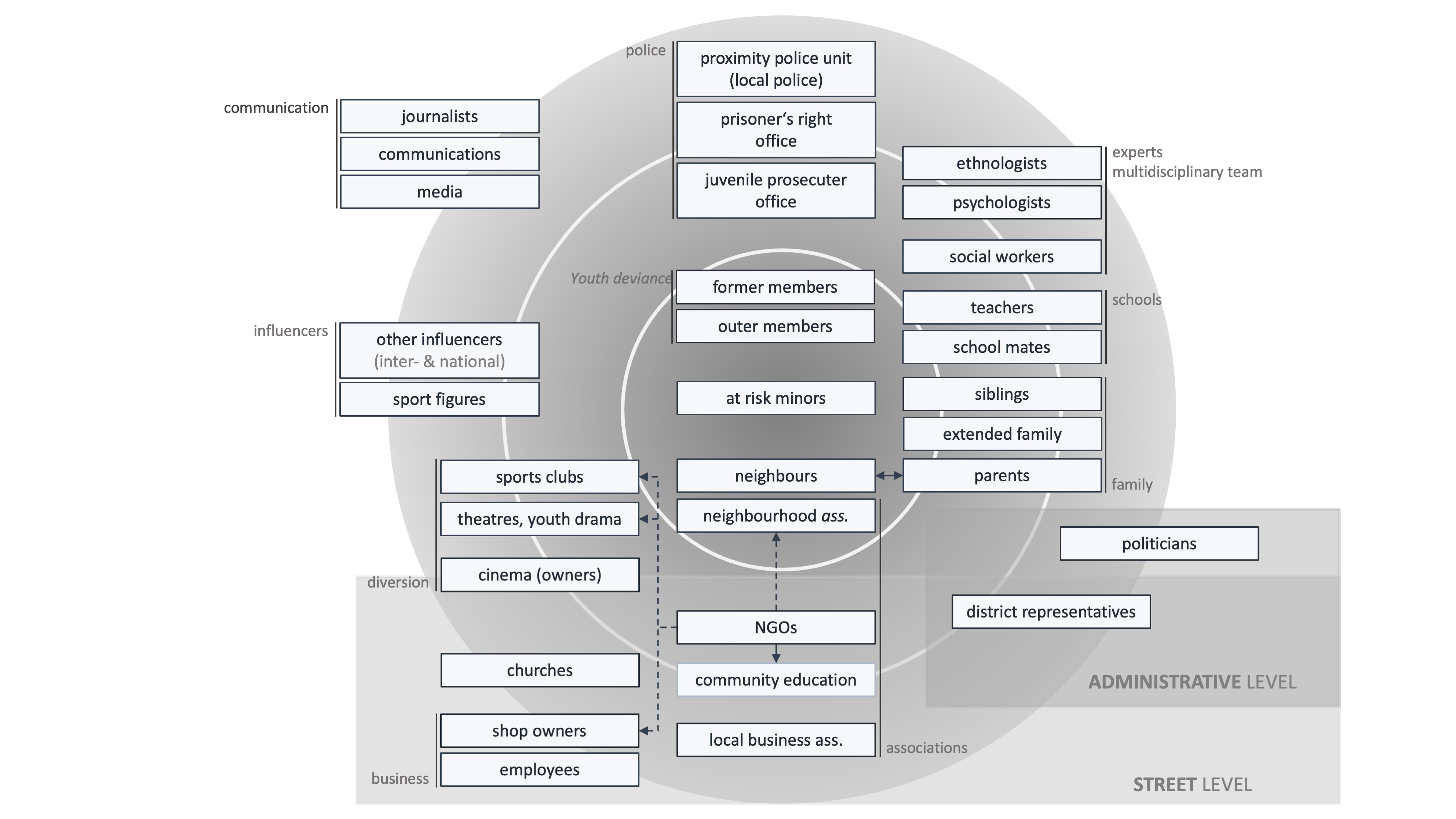}
  \caption{Thematic areas and stakeholder mapping that emerged from the preparatory workshop }\label{fig:stakeholder-mapping}
  \Description{Stakeholder Network for Youth Deviance Prevention. This diagram illustrates the multi-layered stakeholder ecosystem involved in preventing youth deviance. The core represents at-risk minors and their immediate social circle, including family, schoolmates, and neighbors. Surrounding layers include support from local associations, such as neighborhood and business associations, as well as diversion activities (sports clubs, theaters). Additionally, there is involvement from experts, such as psychologists and social workers, law enforcement (local police, juvenile prosecutor's office), and communication channels (journalists, media). Administrative and street levels encompass representatives, NGOs, and various community organizations.}
\end{figure}
%TC:endignore

\section{Detailed surveys
and procedures used in our work}

\subsection{Evaluation Survey Used during the Training of Users (Section~\ref{sec:result-post-train})} \label{sec:appendix-survey-january}

\begin{itemize}
    \item Do you agree with the following statements?\\
    \textit{Each question can be answered with the following Likert scale answers: Completely agree / Rather agree / Rather disagree / Completely disagree / No answer}.

    \begin{enumerate}
        \item I understand how \textit{Sbocciamo Torino} initiative will work.
        \item I understood my role in supporting the implementation of \textit{Sbocciamo Torino}.
        \item I understood the importance of a multilateral approach in the development and implementation of \textit{Sbocciamo Torino}.
        \item The use of the dashboard is innovative for the local government in the city to prevent juvenile delinquency.
        \item To achieve this initiative, it is important to take an inclusive approach that takes into account factors such as gender, age, and disability.
        \item I will use the information and knowledge gained from this training in my work.
        \item I feel prepared to use \textit{Sbocciamo Torino}.
    \end{enumerate}
    \item What of the following best describes your organization?  \\
    \textit{Possible options: Third Sector (NGO, association, cooperative) / Municipal office / Local police / Citizens' association / Politician / School or university / Other (specify)}
\end{itemize}

\subsection{Evaluation Questionnaire Used during the Demonstration (Section~\ref{sec:result-post-demon})} \label{sec:appenix-questionnaire-may}

\begin{enumerate}
    \item Do you consider the goals of \textit{Sbocciamo Torino} initiative to be clear?

    \item What were the main obstacles you encountered in the development of this initiative?

    \item In your opinion, what are the main reasons behind the low interest from associations registered in \textit{Sbocciamo Torino} events?

    \item Are there specific aspects of the initiative that you think could be improved to increase interest and effectiveness?

    \item Do you have any suggestions for encouraging greater participation and collaboration among the parties involved?

    \item For future similar initiatives, in light of \textit{Sbocciamo Torino}, can you indicate in one or two words what the city could improve?

    \item How could we improve communication channels on the progress of the initiative?
    \item What of the following best describes your organization?  \\
    \textit{Possible options: Third Sector (NGO, association, cooperative) / Municipal office / Local police / Citizens' association / Politician / School or university / Other (specify)}
\end{enumerate}

\subsection{Procedure for the committee meeting (Section~\ref{sec:committee-meeting})} \label{sec:appendix-committee-meeting-process}

\begin{enumerate}
    \item \textbf{Introduction (10 minutes)}: Presenting Agenda and Ice-breaking Activities
    
    \item \textbf{Discussion and Co-Design (70 minutes)}:

    \begin{enumerate}
        \item Representatives present and analyze data, outlining the current situation of the different neighborhoods
        
        \item A collaborative data analysis session, brainstorming, and co-design sessions shall follow to produce intervention ideas based on the neighborhoods’ needs as shown through data and experienced by the members.
        
        \item The committee assesses the feasibility of proposed interventions, considering risk factors and cost-benefit analysis, and compares them with similar interventions implemented in the past.
        
        \item After selecting the intervention, the appropriate professional profiles needed to implement the intervention are identified. For example, an intervention to mitigate the phenomenon of the high rate of school dropout in a neighborhood will presumably involve more school workers, educators, and mediators.
        
        \item Together with the sociologist and the data scientist, the committee proposes that the data be collected during the implementation to monitor the trend of the phenomenon and then measure the intervention's impact.
        
    \end{enumerate}

    \item \textbf{Finalizing Interventions (10 minutes)}: Concluding the intervention ideas and policy directions, setting objectives for the next meeting.

    \item \textbf{Miscellaneous (10 minutes)}: Addressing any other matters not on the agenda.

\end{enumerate}

\subsection{Procedure for the extended committee meeting (Section~\ref{sec:extended-committee-meeting})} \label{sec:appendix-extended-committee-meeting-process}

\begin{enumerate}
    \item \textbf{Introduction (5 minutes)}: Presenting the agenda.
    \item \textbf{Intervention Presentation (20 minutes)}: Stakeholders (or the project manager) present the proposed intervention to the councilors.
    \item \textbf{Discussion (20 minutes)}: Review of the intervention and approval of the action plan.
    \item \textbf{Next Steps (15 minutes)}: Discussion of follow-up actions.
\end{enumerate}

\end{document}